\documentclass[review]{elsarticle}
\biboptions{sort&compress} 

\usepackage{epsfig,amsmath,amssymb}
\usepackage{enumerate}
\usepackage{subfigure}
\usepackage{graphicx} 
\usepackage{epstopdf}
\usepackage{booktabs}
\usepackage{multirow}
\usepackage{xcolor}
\usepackage{threeparttable}
\journal{arXiv}

\usepackage{amsmath,amssymb}
\usepackage{subfigure,algorithm,algorithmic}



\usepackage{amsmath}

        \makeatletter
        \def\fps@eqnfloat{!t}
        \def\ftype@eqnfloat{4}
        
        \newenvironment{eqnfloat*}
               {\@dblfloat{eqnfloat}}
               {\end@dblfloat}
        \makeatother
\makeatletter
\let\oldabs\abs
\def\abs{\@ifstar{\oldabs}{\oldabs*}}
\let\oldnorm\norm
\def\norm{\@ifstar{\oldnorm}{\oldnorm*}}
\makeatother

\renewenvironment{thebibliography}[1]{%
	\begin{oldthebibliography}{#1}%
		\setlength{\parskip}{0ex}%
		\setlength{\itemsep}{0ex}%
	}%
	{%
	\end{oldthebibliography}%
}%

\begin{document}
	\begin{frontmatter}
		\title{Time-Range FDA Beampattern Characteristics \tnoteref{t1}}
		\tnotetext[t1]{This work was supported in part by National Natural Science Foundation of China 62171092, in part by Sichuan Science and Technology Program under grant 2018RZ0141, and in part by the Swedish SRA ESSENCE (grant no. 2020 6:2).}
		
		%
		%
		%
		%
		%
		%

		\author[addr]{Wenkai Jia\corref{cor}}
		\address[addr]{School of Information and Communication Engineering, University of Electronic Science and Technology of China, 611731, Chengdu, P. R. China.}
		\ead{wenkai.jia@matstat.lu.se}

		\author[addrr]{Andreas Jakobsson}
		\address[addrr]{Division of Mathematical Statistics,
			Center for Mathematical Sciences, Lund University,  SE-22100, Lund, Sweden. }
		\ead{andreas.jakobsson@matstat.lu.se}
		
		\author[addr]{Wen-Qin Wang}
		\cortext[cor]{Corresponding author.}
		\ead{wqwang@uestc.edu.cn}
		
		%
		%
		
\begin{abstract}
Current literature show that frequency diverse arrays (FDAs) are able of producing range-angle-dependent and time-variant transmit beampatterns, but the resulting time and range dependencies and their
characteristics are still not well understood. 
This paper examines the FDA transmission model and the model for the FDA array factor, considering their time-range relationship.
We develop two FDA transmit beampatterns, both yielding the auto-scanning capability of the FDA transmit beams.
The scan speed, scan volume, and initial mainlobe direction of the beams are also analyzed.
In addition, the equivalent conditions for the FDA integral transmit beampattern and the multiple-input multiple-output (MIMO) beampattern are investigated.
Various numerical simulations illustrate the auto-scanning property of the FDA beampattern and the proposed equivalent relationship with the MIMO beampattern, providing the basis for an improved understanding and design of the FDA transmit beampattern.
\end{abstract}
\end{frontmatter}

\section{Introduction}
The concept of a frequency diverse array (FDA) was first proposed in 2006 \cite{2006Frequency}, and was patented by Wicks and Antonik in \cite{2008Frequency}.
As compared to a phased-array, FDA employs different carrier frequencies across the array elements to generate a transmit beampattern which is a function of the range, angle, and time \cite{2008Frequenc,2009Developments,jia2023waveform}. 
This interesting fusion of frequency and spatial diversities provides FDA with broad application prospects and the topic has attracted notable interest since \cite{wang2016moving,9161264,9440819,9944910}.
As the standard FDA usually produces an S-shaped beampattern in the range-angle plane \cite{jia2023joint}, logarithmic \cite{6951408}, time-modulated \cite{2014Frequency,2016Range}, random \cite{2017The}, and optimized \cite{2016Frequency} frequency offsets (FOs) have been suggested to decouple the beampattern.
Combining co-located multiple-input multiple-output (MIMO) with waveform diversity \cite{li2007mimo}, distributed MIMO with spatial diversity \cite{haimovich2007mimo}, a focused transmit beampattern has also been synthesized by optimizing the transmitted baseband waveform \cite{2013Frequency}. Furthermore, to obtain the desired dot-shaped time-invariant beampattern, the alternating direction method of multipliers (ADMM) algorithm was adopted to design the transmitted weight matrix in \cite{2021Transmit}. On the other hand, multiple methods have been proposed to explore the range-angle-dependent property of the FDA transmit beampattern, such as dual-pulse FDA \cite{2014Rge}, FDA-MIMO \cite{2014Transmit}, and FDA subarrays \cite{2014Subarray}.
Basit {\em et al} \cite{2019Rnge} studied the use of FDA with nonlinearly progressive FOs for joint radar and
communication functionalities.
Xu \emph{et al.} proposed an FDA strategy for identifying and suppressing deceptive mainlobe interference and for suppressing range-ambiguous clutter \cite{2015Deceptive}, which was shown to yield better performance than that of a phased-array \cite{7181636}. Additionally, the use of FDA for low probability of intercept (LPI) transmit beamforming \cite{2016Overview,2017Cognitive}, secure communications \cite{2016Artificial}, and target detection \cite{9212375} have been reported.

Since the angle-time coupling is the aspect of an FDA antenna that differs it from a conventional phased-array or MIMO systems, it is more meaningful to effectively utilize the time variance to improve the performance than to strive to decouple the beampattern.
It is not only that the FDA antenna's time-variant property cannot be adequately suppressed, but the obtained pseudo-static or focused FDA would lose its advantage in comparison to phased-array and MIMO antennas.
The time variance of the FDA transmit beampattern was analyzed in \cite{2020Enhanced} and \cite{8074796}, but the time-range relations are still not sufficiently detailed.

In this paper, considering its time-range dependencies, an accurate model of the FDA instantaneous transmit beampattern (FITB) is formulated and then used to detail the auto-scanning capability of an FDA transmit beam.
Auto-scanning here refers to autonomously and continuously changing the direction of the mainlobe of the beam within the pulse duration without electronically adjusting the phase shifters.
The scan speed, scan volume, and initial mainlobe direction of the beam are also detailed. Comparing the resulting beampattern with that of a phased-array, it can be seen that the uniform linear FO is the optimal choice for FDA.
In addition, we present the FDA integral transmit beampattern model (FGTB), which provides a method to calculate the amount of energy radiated by the FDA to a certain azimuth angle during the pulse duration. The relation between the FGTB and the MIMO beampatterns are also investigated, revealing the conditions for equivalence for these systems.

The remaining sections are organized as follows.
Section \ref{Sec2} reformulate the FDA transmitted electric field model considering the underlying time-range dependencies.
In Section \ref{Sec3}, 
the proposed FITB is derived and its scan speed, scan volume, and initial mainlobe direction are derived.
In Section \ref{Sec4}, the influence of the FO on the proposed FITB and the relationship between the FDA and a phased-array are discussed.
Next, the FGTB model is derived in Section \ref{Sec5}, followed by a comparison with the MIMO beampattern.
Finally, conclusions are drawn in Section \ref{Sec6}.

\section{Signal Model}\label{Sec2}

\begin{figure}[t]
	\centering
	\includegraphics[width=0.6\textwidth]{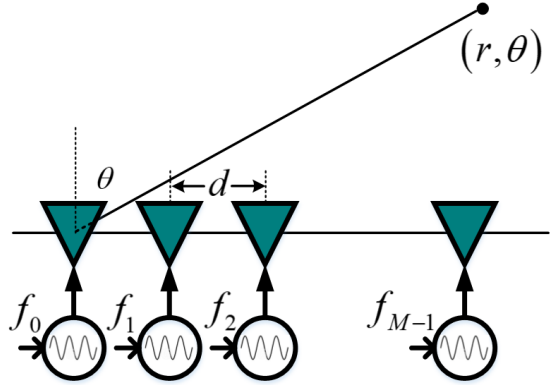}
	\caption{An FDA transmitter configuration.}\label{fig1}
\end{figure}

The transmitted electric field at a given far-field point target for a range-angle pair $\left( {{r},\theta } \right)$ with the target slant range ${{r}}$, using the first element as the reference, is the sum of the field components radiated by the individual FDA elements, as illustrated in Fig. \ref{fig1}, such that \cite{7740083}
\begin{equation}
	{{E}_{T}}\left( r,\theta ,t \right)=\sum\limits_{m=0}^{M-1}{\frac{w_{m}^{c}}{{{r}_{m}}}e\left( \theta \left| {{f}_{m}} \right. \right)\phi \left( t-\frac{{{r}_{m}}}{c} \right){{e}^{j2\pi {{f}_{m}}\left( t-\frac{{{r}_{m}}}{c} \right)}}},
\end{equation}
for $t\in \left[ \frac{{{r}}}{c},\frac{{{r}}}{c}+{{T}_{p}} \right]$,
where ${w_m},m=1,2,...,M-1$ is the beamformer weight of the $m$-th element with $M$ denoting the number of transmit elements. Furthermore,
$e\left( \theta \left| {{f}_{m}} \right. \right)$ represents the radiation pattern of the $m$-th element evaluated at
its carrier frequency $f_m$, ${{r}_{m}}=r-md\sin \theta$ denotes the slant range from the $m$-th element to the target, the $r_{m}^{-1}$ factor accounts for the free space loss, and $\phi \left( t \right)$ is the baseband complex waveform with unit energy, i.e.,
$\int_{{T_p}} {\phi \left( t \right){\phi^c}\left(t\right)dt}  = 1$, where ${T_p}$ and $(\cdot)^c$ are the pulse duration and the conjugate operators, respectively.
It is reasonable to assume that the radiation patterns are nearly the same within the FDA's transmission bandwidth, i.e., $e\left( \theta \left| {{f}_{m}} \right. \right)\approx e\left( \theta \left| {{f}_{c}} \right. \right)$, where $f_c$ is the center carrier frequency.
In addition, under far-field conditions, since ${{r}_{m}}\gg \left( M-1 \right)d\sin \theta$, it may be assumed that the distance attenuation has an identical influence on the electric field amplitudes of all array elements. Therefore, the FDA transmitted electric field can be well modeled as
\begin{equation}
	{{E}_{T}}\left( r,\theta ,t \right)=\sum\limits_{m=0}^{M-1}{\frac{w_{m}^{c}}{r}e\left( \theta \left| {{f}_{c}} \right. \right)\phi \left( t-\frac{{{r}_{m}}}{c} \right){{e}^{j2\pi {{f}_{m}}\left( t-\frac{{{r}_{m}}}{c} \right)}}},
\end{equation}
for $t\in \left[ \frac{{{r}}}{c},\frac{{{r}}}{c}+{{T}_{p}} \right]$, where $\phi \left( {t - \frac{{{r_m}}}{c}} \right) \approx \phi \left( {t - \frac{{{r}}}{c}} \right)$ based on the narrowband assumption
\begin{equation}\label{eq3}
	\frac{Md}{c}\ll \frac{1}{{{B}_{\phi \left( t \right)}}}
\end{equation}
with ${B_{\phi \left( t \right)}}$ being the bandwidth of the baseband envelope $\phi \left( t \right)$. 
In order to admit to a closed-form representation, uniform linear frequency increments are adopted (the effect of nonlinear FO on the beampattern will be analyzed in Section \ref{Sec4}, i.e., 
\begin{equation}
	{f_m} = {f_c} + m\Delta f,
\end{equation}
for $m = 0,1,...,M - 1$, where $\Delta f$ denotes the FO.
Then, the FDA transmitted electric field can be reformulated as
\begin{equation}\label{eq5}
	\begin{aligned}
		& {{E}_{T}}\left( r,\theta ,t \right)=\frac{1}{r}e\left( \theta \left| {{f}_{c}} \right. \right)\phi \left( t-\frac{r}{c} \right){{e}^{j2\pi {{f}_{c}}\left( t-\frac{r}{c} \right)}} \\ 
		& \kern 25pt \times \underbrace{{{\mathbf{w}}^{H}}\left[ {{\mathbf{a}}_{T}}\left( \Delta f,t,r \right)\odot {{\mathbf{a}}_{T}}\left( \theta  \right)\odot {{\mathbf{a}}_{T}}\left( \Delta f,\theta  \right) \right]}_{\text{A}rray \kern 2pt factor} \\ 
	\end{aligned}
\end{equation}
for $t \in \left[ {\frac{{{r}}}{c},\frac{{{r}}}{c} + {T_p}} \right]$, where $(\cdot)^{T}$, $(\cdot)^{H}$, and $\odot$ denote the
transpose, conjugate transpose, and Hadamard operators, respectively, and
\begin{subequations}
	\begin{equation}
		\kern -40pt \mathbf{w}={{\left[ \begin{matrix}
					{{w}_{0}} & {{w}_{1}} & ... & {{w}_{M-1}}  \\
				\end{matrix} \right]}^{T}}
	\end{equation}
	\begin{equation}
		{{\mathbf{a}}_{T}}\left( \Delta f,t,r \right)={{\left[ \begin{matrix}
					1 & {{e}^{j2\pi \Delta f\left( t-\frac{r}{c} \right)}} & ... & {{e}^{j2\pi \left( M-1 \right)\Delta f\left( t-\frac{r}{c} \right)}}  \\
				\end{matrix} \right]}^{T}}
	\end{equation}
	\begin{equation}
		\kern 20pt {{\mathbf{a}}_{T}}\left( \theta  \right)={{\left[ \begin{matrix}
					1 & {{e}^{j2\pi \frac{{{f}_{c}}}{c}d\sin \theta }} & ... & {{e}^{j2\pi \frac{{{f}_{c}}}{c}\left( M-1 \right)d\sin \theta }}  \\
				\end{matrix} \right]}^{T}}
	\end{equation}
	\begin{equation}
		\kern 5pt {{\mathbf{a}}_{T}}\left( \Delta f,\theta  \right)={{\left[ \begin{matrix}
					1 & {{e}^{j2\pi \Delta f\frac{d\sin \theta }{c}}} & .. & {{e}^{j2\pi \Delta f\frac{{{\left( M-1 \right)}^{2}}d\sin \theta }{c}}}  \\
				\end{matrix} \right]}^{T}}.
	\end{equation}
\end{subequations}
Using uniform weighting, i.e., $\mathbf{w}={{\mathbf{1}}_{M}}$, the FDA array factor can be rewritten as \cite{wang2013range,2016Frequency,2016Fruency,2021Transmit}
\begin{equation}\label{eq7}
	\begin{aligned}
		& {\mathcal{A}_{old}}\left(r,\theta, t  \right)=\left| {{\mathbf{1}}^{H}}\left[ {{\mathbf{a}}_{T}}\left( \Delta f,t,r \right)\odot {{\mathbf{a}}_{T}}\left( \theta  \right)\odot {{\mathbf{a}}_{T}}\left( \Delta f,\theta  \right) \right] \right| \\ 
		& \kern 10pt \approx \left| \frac{\sin \left[ M\pi \left( \Delta ft-\frac{\Delta fr}{c}+\frac{{{f}_{c}}d\sin \theta }{c}+\frac{\Delta fd\sin \theta }{c} \right) \right]}{\sin \left[ \pi \left( \Delta ft-\frac{\Delta fr}{c}+\frac{{{f}_{c}}d\sin \theta }{c}+\frac{\Delta fd\sin \theta }{c} \right) \right]} \right| \\ 
	\end{aligned}
\end{equation}
where the approximation
\begin{equation}
	{e^{j2\pi {m^2}\Delta f\frac{{d\sin \theta }}{c}}} \approx {e^{j2\pi m\Delta f\frac{{d\sin \theta }}{c}}},
\end{equation}
for $m = 0,1,...,M - 1$, has been used.
One observes from \eqref{eq7} that, different from phased-arrays, the obtained array factor features an additional dependence on the range and the time as well as the angle, a property which renders the field of an FDA different from that of a phased-array.
However, the range of the time variable $t$, $t\in \left[ \frac{r}{c},\frac{r}{c}+{{T}_{p}} \right]$, is inadvertently ignored in \eqref{eq7}.
In order to fully explore the characteristics of the beampattern, one may use the variable substitution $t' = t - \frac{{{r}}}{c}$, yielding
\begin{equation}\label{eq9}
	\mathcal{A}\left( t',\theta  \right)\approx \left| \frac{\sin \left[ M\Upsilon \left( t',\theta  \right) \right]}{\sin \left[ \Upsilon \left( t',\theta  \right) \right]} \right|
\end{equation}
where
\begin{equation}
	\Upsilon \left( t',\theta  \right)=\pi \left[ \Delta ft'+\left( {{f}_{c}}+\Delta f \right)\frac{d\sin \theta }{c} \right],
\end{equation}
for ${t}'\in \left[ 0,{{T}_{p}} \right]$.
It must be emphasized that the time uniqueness should be guaranteed. For a given target distance, the trip delay is constant.
It may be noted from \eqref{eq9} that the new array factor formulation is time-angle-dependent within the pulse duration, which also exhibits a considerable difference from the static beampattern (array factor) of a conventional phased-array.
As the array factor is time-variant, we here term this the FDA instantaneous transmit beampattern (FITB). However, it is worth noting that the range-dependent property proposed in \cite{2016Frequency,2016Fruency,2021Transmit} will not hold for the FDA transmit beampattern.

Current research on the FDA transmit beampattern mainly focuses on the static range-angle-dependent beampattern by fixing the time variable $t$ at the instantaneous moment $t=0$ in \eqref{eq7}, see, e.g., \cite{2014Frequency,2016Range,2017Sutions}.
As alternatives, the focused transmit beampatterns in the range-angle plane has also been synthesized using designed logarithmic \cite{6951408}, time-modulated \cite{2014Frequency,2016Range}, random \cite{2017The}, or optimized \cite{2016Frequency} FOs.
However, the range-dependent theory of the FDA beampatterns was challenged by the consideration of the Huygens-Fresnel principle \cite{urone1998college}.  
Since the time variable $t$ belongs to $t \in \left[ {\frac{{{r}}}{c}, \frac{{{r}}}{c} + {T_p}} \right]$, the conclusion of a range-dependent array factor is not convincing, This in turn implies that the characteristics of the FDA transmit beampattern are still in need of further clarification. 

\section{the characteristics of FITB}\label{Sec3}
\subsection{Zero-time Cut}
\begin{figure}[t]
	\centering
	\includegraphics[width=0.6\textwidth]{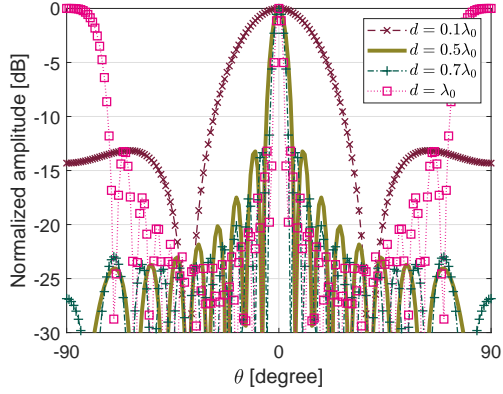}
	\caption{Zero-time cut of the FITB.}\label{fig2}
\end{figure}

Although the proposed FITB is time-variant, it is still meaningful to analyze its characteristics at a given time instant.
When $t' = 0$ in \eqref{eq9}, the zero-time cut of FITB can be expressed as
\begin{equation}
	\mathcal{A}\left( {0,\theta } \right) \approx \left| {\frac{{\sin \left[ {M\pi \left( {\frac{{{f_c}d\sin \theta }}{c} + \frac{{\Delta fd\sin \theta }}{c}} \right)} \right]}}{{\sin \left[ {\pi \left( {\frac{{{f_c}d\sin \theta }}{c} + \frac{{\Delta fd\sin \theta }}{c}} \right)} \right]}}} \right|
\end{equation}
where the zero-time cut refers to the beampattern at the instantaneous moment $t' = 0$.
This can be understood as the electric field radiated by the $M$ FDA antenna elements combine in phase at this time. 
Then, the first null beamwidth, i.e., the mainbeam width, is
\begin{equation}
	{{\Omega }_{0}}  = {\sin ^{ - 1}}\left( {\frac{c}{{{f_c}Md}}} \right),
\end{equation}
with its peak being located at
\begin{equation}
	{{\theta }_{0,peak}} = {\sin ^{ - 1}}\left( {k\frac{c}{{{f_c}{d}}}} \right),\kern 15pt k \in \mathbb{Z}
\end{equation}
where $\mathbb{Z}$ denote the integer set.
Fig. \ref{fig2} shows the zero-time cut of the FITB, using $M = 16$, ${f_c} = 10 \kern 2pt GHz$, and $\Delta f = 100{\kern 2pt}Hz$. It may be seen that the larger the inter-element spacing $d$, the narrower the beamwidth is. If $d$ exceeds half the reference wavelength ${\lambda _0} = \frac{c}{{{f_c} + M\Delta f}}$, the grating lobes will appear around $ \pm {90^o}$.

\subsection{Auto-scanning}\label{Sec3b}
According to \eqref{eq9}, the FITB has peaks at locations satisfying
\begin{equation}
	\Delta ft' + \frac{{{f_c}d\sin \theta }}{c} + \frac{{\Delta fd\sin \theta }}{c} = k,
\end{equation}
for ${t}'\in \left[ 0,{{T}_{p}} \right]$ and $k \in \mathbb{Z}$.
Therefore, the mainbeam at time instant $t'$ will be equivalently steered to direction
\begin{equation}\label{eq15}
	{{\theta }_{t}}  = {\sin ^{ - 1}}\left( {\frac{{ck - c\Delta ft}}{{{f_c}d + \Delta fd}}} \right).
\end{equation}
Taking the derivative of both sides of \eqref{eq15} with respect to $t'$, the spatial scan speed of FITB can be obtained as
\begin{equation}\label{eq16}
	\begin{aligned}
		\frac{d\theta }{d{t}'}&=-\frac{c\Delta f}{\left( {{f}_{c}}+\Delta f \right)d\cos \theta } \\ 
		& \kern -3pt \overset{d=\frac{{{\lambda }_{0}}}{2}}{\mathop{=}}\,-\frac{2\Delta f\left[ {{f}_{c}}+\left( M-1 \right)\Delta f \right]}{\left( {{f}_{c}}+\Delta f \right)\cos \theta } \\ 
		&\kern 1pt \approx -\frac{2\Delta f}{\cos \theta }. \\ 
	\end{aligned}
\end{equation}
Furthermore, the beamwidth or resolution ${{\Theta }_{{{t}'}}}$ of the field at a given time instant $t'$ satisfies 
\begin{equation}\label{eq17}
	\begin{aligned}
		\Upsilon \left( {t}',\theta  \right)+\Delta {{\Upsilon }_{{{t}'}}}&=\pi \left[ \Delta f{t}'+\left( {{f}_{c}}+\Delta f \right)\frac{d\left( \sin \theta +{{\Theta }_{{{t}'}}} \right)}{c} \right] \\ 
		& =\Upsilon \left( {t}',\theta  \right)+\pi \left( {{f}_{c}}+\Delta f \right)\frac{d{{\Theta }_{{{t}'}}}}{c}. \\ 
	\end{aligned}
\end{equation}
Adopting the convenient 4-dB Rayleigh width criterion, selecting
$\Delta {{\Upsilon }_{{{t}'}}} =\frac{\pi }{N}$ in \eqref{eq17}, yields
\begin{equation}
	{{\Theta }_{{{t}'}}}=\Theta =\frac{c}{M\left( {{f}_{c}}+\Delta f \right)d}\approx \sin \left( {{\theta }_{0,peak}} \right)\approx \frac{{{\lambda }_{0}}}{Md}.
\end{equation}
It may be noted that ${{\Theta }}$ relates to the effective angular extension of the field, being linked to the azimuth resolution $\bar{\theta }$ as
\begin{equation}
	\begin{aligned}
		\frac{{{\lambda }_{0}}}{Md}&=\sin {{\theta }_{4dB+}}-\sin {{\theta }_{4dB-}} \\ 
		& =2\cos \frac{{{\theta }_{4dB+}}+{{\theta }_{4dB-}}}{2}\cdot \sin \frac{{{\theta }_{4dB+}}-{{\theta }_{4dB-}}}{2} \\ 
		& =2\cos \frac{{{\theta }_{4dB+}}+{{\theta }_{4dB-}}}{2}\cdot \sin \frac{\Delta \theta }{2} \\ 
		& \approx \bar{\theta } \cdot \cos {{\theta }_{o,peak}} \\ 
	\end{aligned}
\end{equation}
implying that 
\begin{equation}\label{eq20}
	\bar{\theta }=\frac{{{\lambda }_{0}}}{Md\cos {{\theta }_{o,peak}}}.
\end{equation}

From \eqref{eq16} and \eqref{eq20}, one may conclude that
\begin{itemize}
	\item[1)] 
	Since a delay increment along the azimuth dimension does not translate to a relative phase difference for identical sources, evidenced by the lack of any radial gradients along the time axis for $\Delta f=0$, a phased-array can only control the direction of the mainlobe by electronically adjusting the phase shifters. The generated field is therefore static, exhibiting maxima (minima) when the radiated element fields add constructively (destructively).
	However, due to the existence of FO, the phase progression between the FDA elements grows in time, such that the auto-scanning behavior occurs without any phase shifters.
	Since $\cos \theta$ is a decreasing function, the scan speed is nonlinear, directly proportional to the FO and the current azimuth angle.
	This auto-scanning ability allows the beam to spend more time at boresite and less at extreme angles, which can be quite useful in practical applications.
	\item[2)] The resolution of the beam is time-invariant. Within the pulse duration, the beam coverage (scan volume) is
	\begin{equation}
		\begin{aligned}
			{{{\bar{\Theta }}}_{{{T}_{p}}}}&=\frac{c\Delta f{{T}_{p}}}{\left( {{f}_{c}}+\Delta f \right)d} \\ 
			& =\frac{\Delta f{{T}_{p}}2\left[ {{f}_{c}}+\left( M-1 \right)\Delta f \right]}{\left( {{f}_{c}}+\Delta f \right)} \\ 
			& \kern -5pt \overset{\Delta f\ll {{f}_{c}}}{\mathop{\approx }}\,2\Delta f{{T}_{p}} \\ 
		\end{aligned}
	\end{equation}
	which is thus dependent on the FO and the pulse duration $T_p$. In particular, if $\Delta f = \frac{1}{{{T_p}}}$, the FDA can scan through the whole visible azimuth sector within one pulse duration. As a result, the presence of a small FO enables the FDA to transmit automatically scanning beams.
\end{itemize}

\subsection{Initial Mainlobe Direction}\label{Sec3c}
For phased arrays, the transmit weight vector ${\mathbf{w}}$ ensures that the transmit beam is steered in the desired direction.
However, due to the auto-scanning capability of the FDA beam,
the vector will mainly affect the initial mainlobe direction.
For example, setting ${\mathbf{w}}$ as 
\begin{equation}\label{eq22}
	\mathbf{w}={{\left[ {{\mathbf{a}}_{T}}\left( {{\theta }_{0}} \right)\odot{{\mathbf{a}}_{T}}\left( \Delta f,{{\theta }_{0}} \right) \right]}^{c}}
\end{equation}
and substituting \eqref{eq22} into \eqref{eq5} yields
\begin{equation}
	\begin{aligned}
		& {{E}_{T,\mathbf{w}}}\left( r,\theta ,t' \right)\approx \frac{1}{r}e\left( \theta \left| {{f}_{c}} \right. \right)\phi \left( {{t}'} \right){{e}^{j2\pi {{f}_{c}}{t}'}} \\ 
		& \times \underbrace{\sum\limits_{m=0}^{M-1}{{{e}^{-j2\pi m\Delta f{t}'}}{{e}^{-j2\pi \left( {{f}_{c}}+m\Delta f \right)md\frac{\sin \theta -\sin {{\theta }_{0}}}{c}}}}}_{Array \kern 2pt factor}. \\ 
	\end{aligned}
\end{equation}
Thus, the weight-dependent FITB at $t'=0$ can be expressed as
\begin{equation}
	{\mathcal{A}_{\mathbf{w}}}\left( 0,\theta  \right)=\left| \frac{\sin \left\{ M\pi \left[ \frac{\left( {{f}_{c}}+\Delta f \right)d\left( \sin \theta -\sin {{\theta }_{0}} \right)}{c} \right] \right\}}{\sin \left\{ \pi \left[ \frac{\left( {{f}_{c}}+\Delta f \right)d\left( \sin \theta -\sin {{\theta }_{0}} \right)}{c} \right] \right\}} \right|
\end{equation}
with its peak appearing at $\theta  = {\theta _0}$.

\subsection{Numerical Simulation}\label{Sec3d}
In this section, various numerical simulations are presented to evaluate the auto-scanning behavior of the FDA transmitted beampattern.
For a given far field distance, we first verify the relationship between the FO and the beam scan volume.
Then, we examine the angular extension of the beam over different time periods.
Next, we illustrate the effect of the transmit weight vector on the initial mainlobe direction of the beam.
Finally, to explain the difference between the range dependent characteristics used in \cite{2014Frequency,2014Rge,2016Frequency,2017Joint,2017The} and that of the auto-scanning property proposed in this paper, these two beampatterns are compared, at two different far-field distances.
In all simulations, we assumed that the number of FDA transmit array elements is $M = 16$, the reference carrier frequency is $f_c = 10{\kern 2pt}GHz$, and that the inter-element spacing is $d = \frac{{{\lambda _0}}}{2}$ (for a phased-array this implies $d={c}/{2{{f}_{c}}}\;$). For simplicity, a rectangular pulse with pulse duration ${T_p} = 5\kern 2pt \mu s$ is considered as the baseband waveform.

\subsubsection{Scan volume}
\begin{figure}[t]
	\centering  
	\subfigure[$\Delta f = 10{\kern 2pt}kHz$.]{
		\label{fig3a}
		\includegraphics[width=0.45\textwidth]{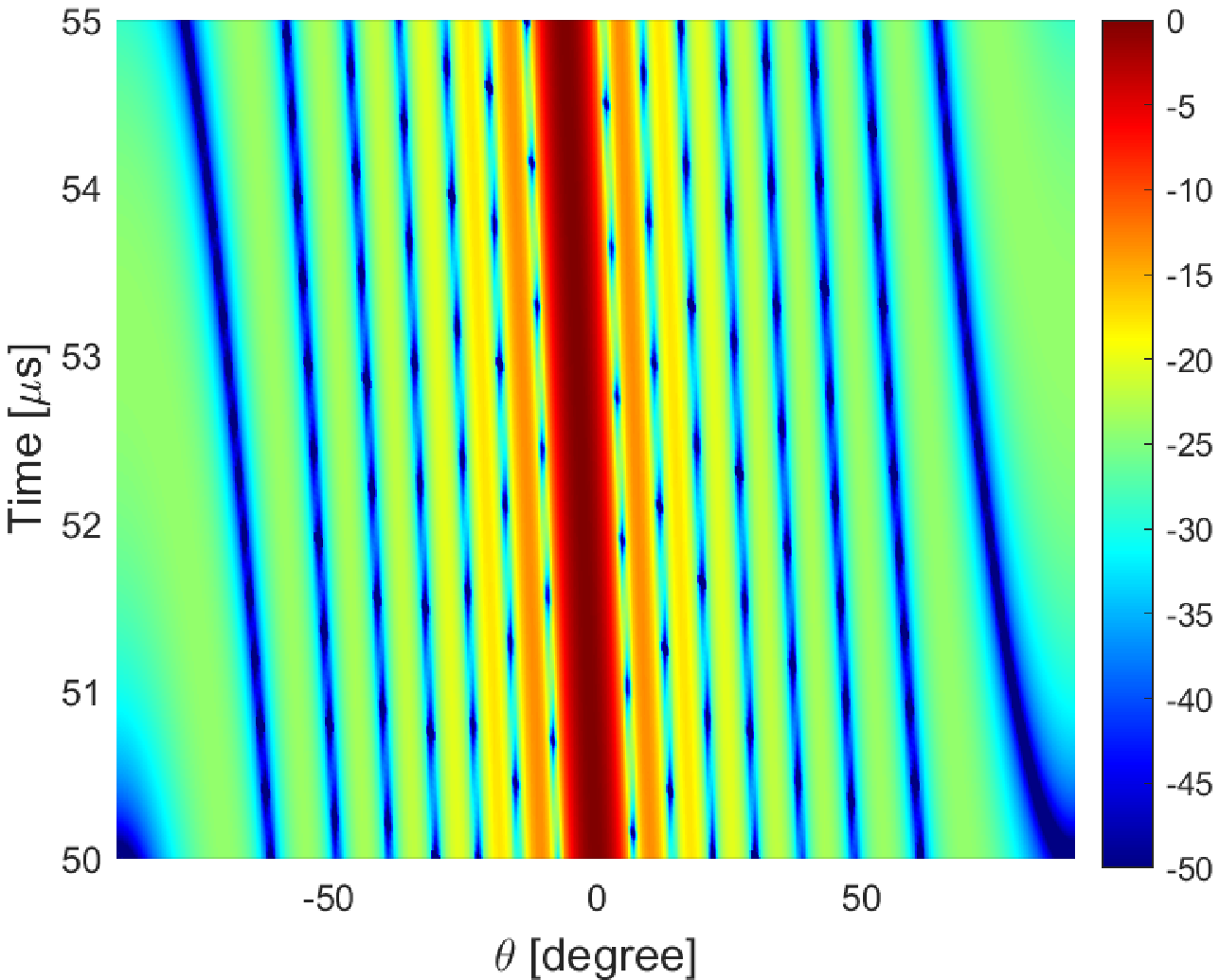}}
	\subfigure[$\Delta f = 30{\kern 2pt}kHz$.]{
		\label{fig3b}
		\includegraphics[width=0.45\textwidth]{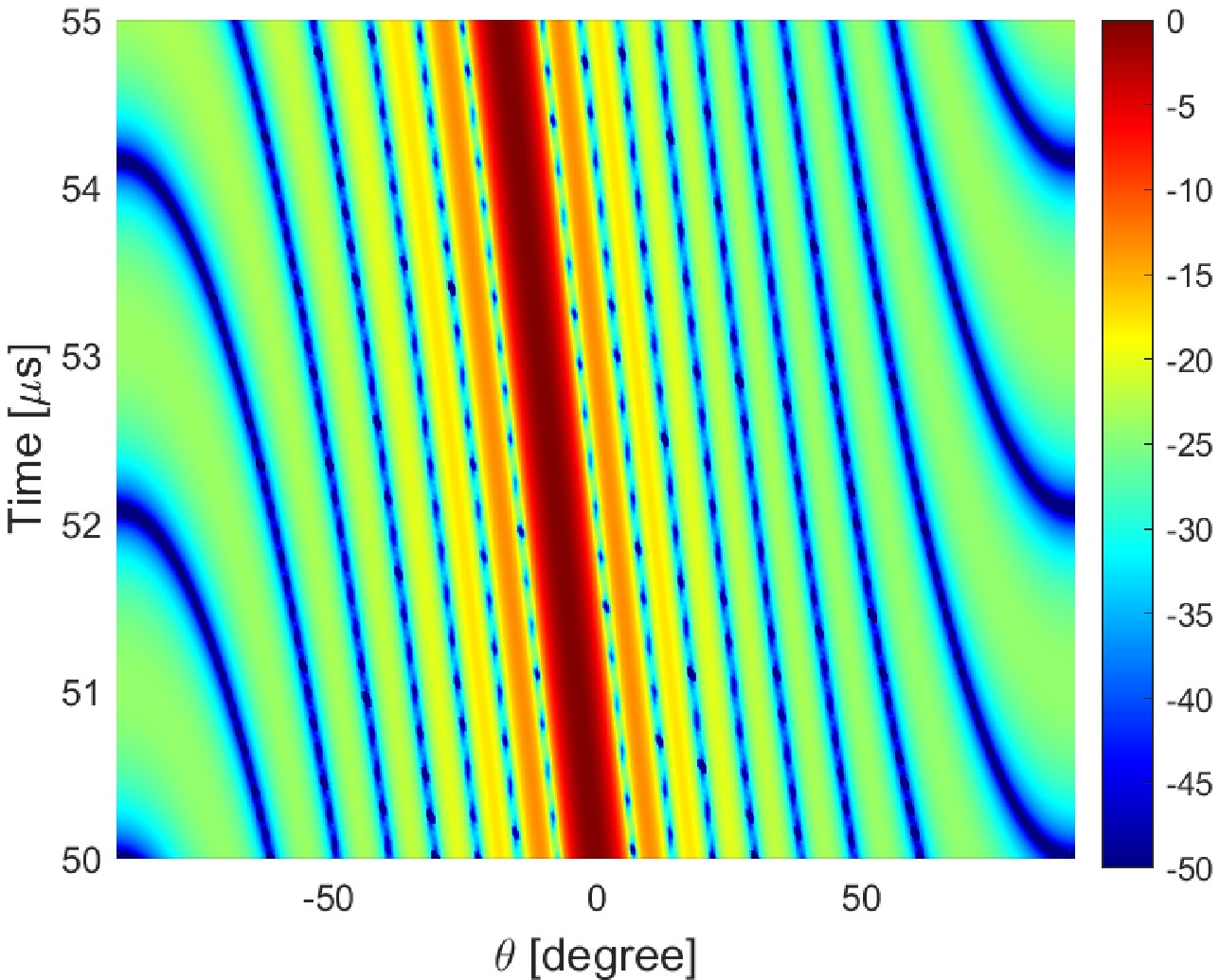}}
	\subfigure[$\Delta f = 200{\kern 2pt}kHz$.]{
		\label{fig3c}
		\includegraphics[width=0.45\textwidth]{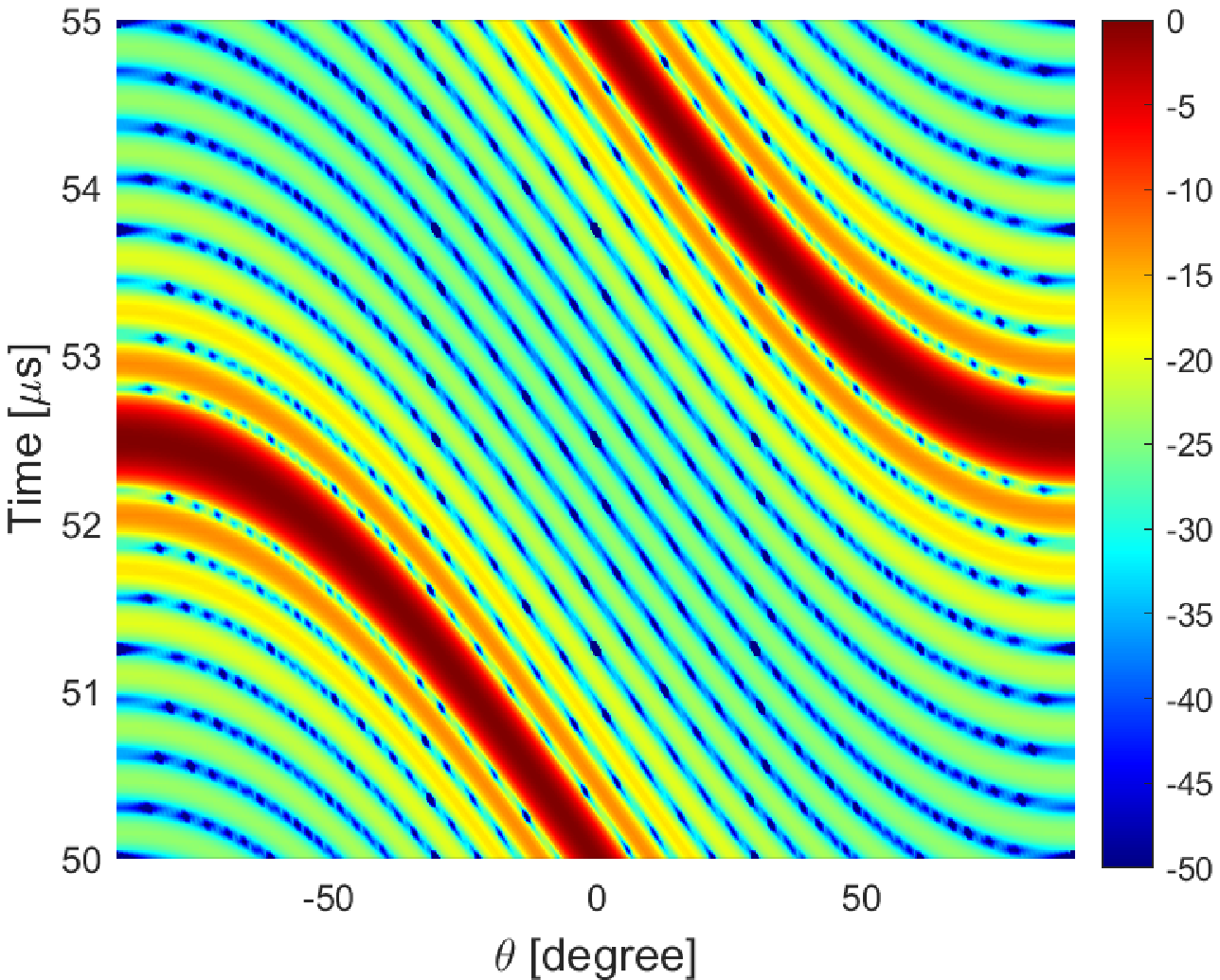}}
	\subfigure[$\Delta f = 400{\kern 2pt}kHz$.]{
		\label{fig3d}
		\includegraphics[width=0.45\textwidth]{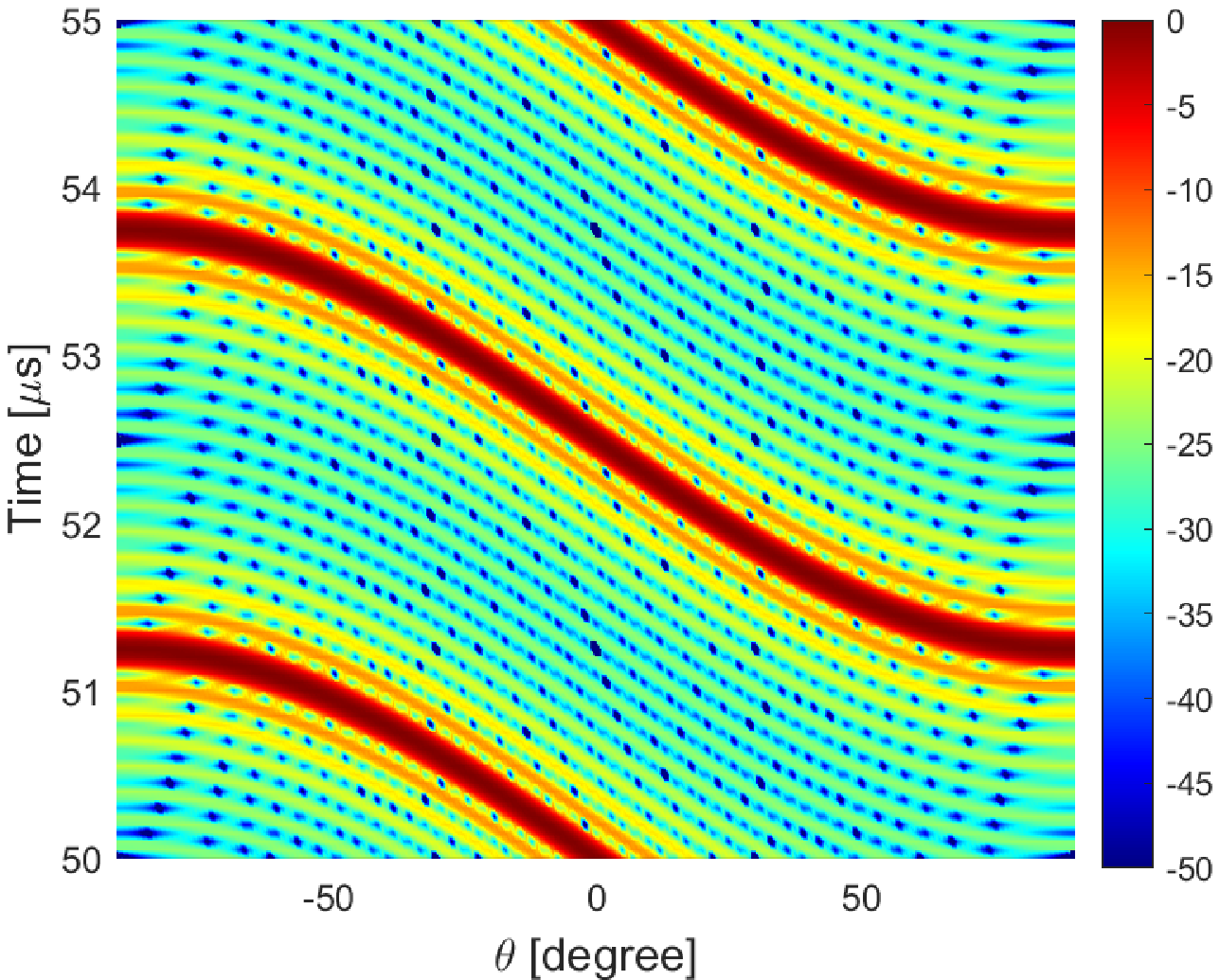}}
	\subfigure[$\Delta f = 0{\kern 2pt}kHz.$]{
		\label{fig3e}
		\includegraphics[width=0.6\textwidth]{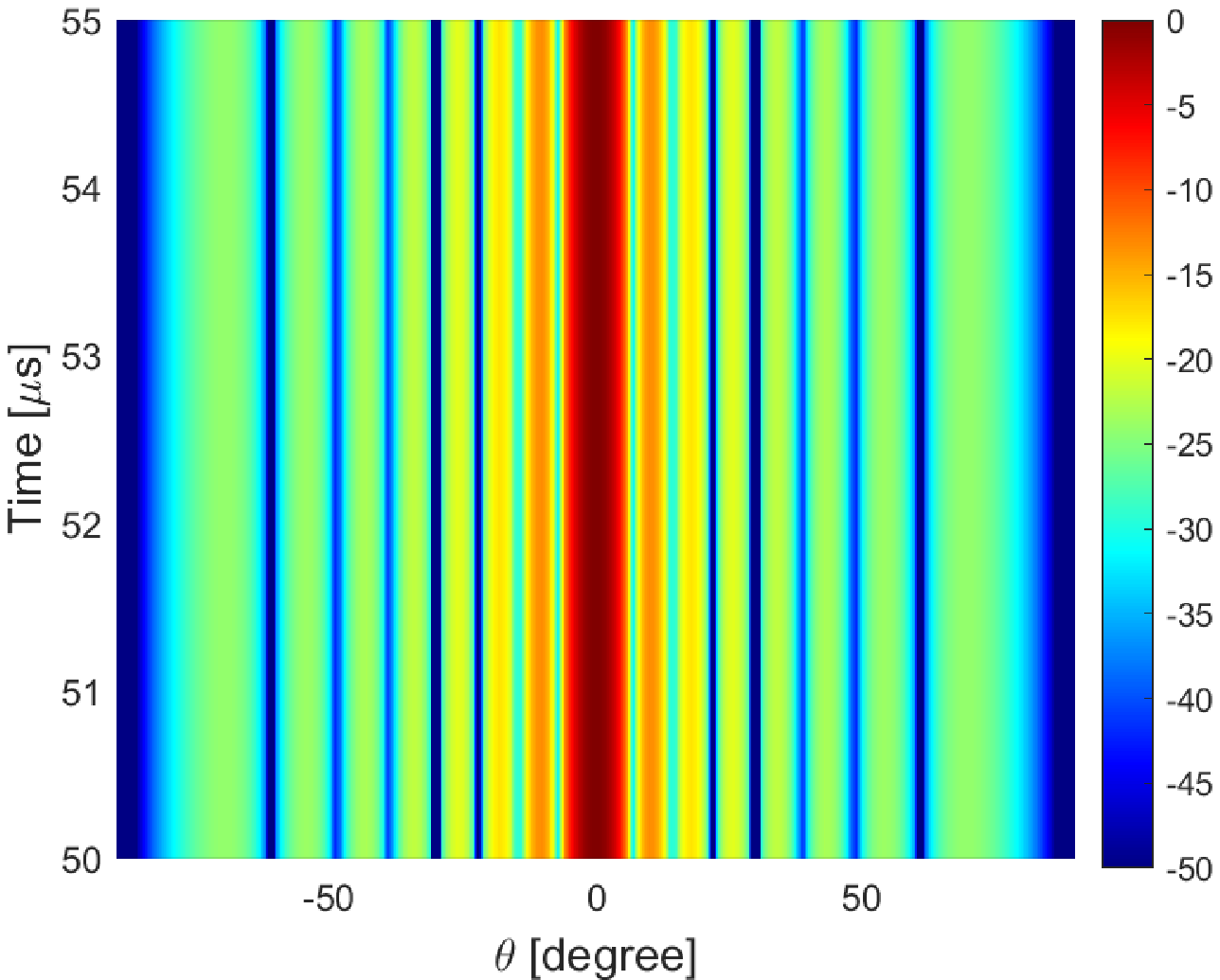}}
	\caption{Electric field emanated from the FDA with different FO configurations at a far-field distance of $r=15 \kern2pt km$. }
	\label{fig3}
\end{figure}

Fig. \ref{fig3} illustrates the simulated FDA electric fields for different FO configurations using a uniform weight vector, at a given far-field distance $r=15 \kern2pt km$, with the field of the phased-array ($\Delta f = 0{\kern 2pt}Hz$) acting as a benchmark.
It can be seen that for a phased-array, the field is static, i.e., it is not spatially scanned over time.
Unlike for a phased-array, the field formed by the FDA is dynamic.
Small FO produces limited beam coverage, almost the same as the phased-array field, although the scan volume becomes larger as the FO increases.
The results also show that the FDA can scan the entire visible azimuth sector within one pulse duration when the FO satisfies $\Delta f=\frac{1}{{{T}_{p}}}$, corresponding to the results in Fig. \ref{fig3c}.
This observation agrees with the results discussed in Section \ref{Sec3b}.
It can also be found that if the FO is larger than $\frac{1}{{{T}_{p}}}$, the FDA is sufficient to scan through the entire visible azimuth sector multiple times within one pulse period.
As shown in Fig. \ref{fig3d}, for this example, the beam scans the entire sector twice.
It is worth reiterating that the auto-scanning behavior of the FDA occurs without any phase shifters.
However, as compared to the phased-array, the radiated energy is also dispersed in the entire observed airspace, which will inevitably lead to a loss of transmission gain.

\subsubsection{Scan speed}
\begin{figure}[t]
	\centering  
	\subfigure[$t\in \left( 50,51.5 \right)\mu s$.]{
		\label{fig4a}
		\includegraphics[width=0.45\textwidth]{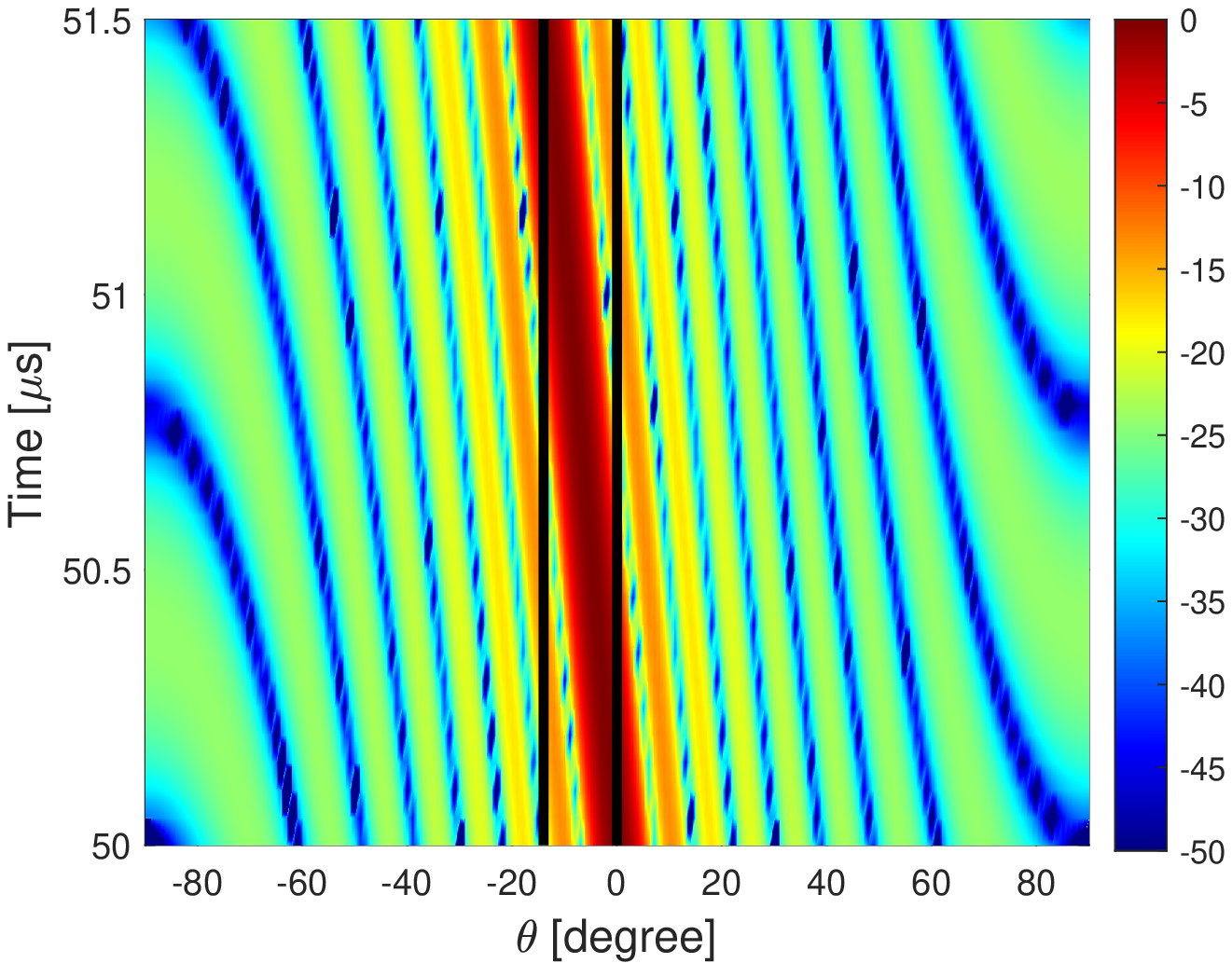}}
	\subfigure[$t\in \left( 53.5,55 \right)\mu s$.]{
		\label{fig4b}
		\includegraphics[width=0.45\textwidth]{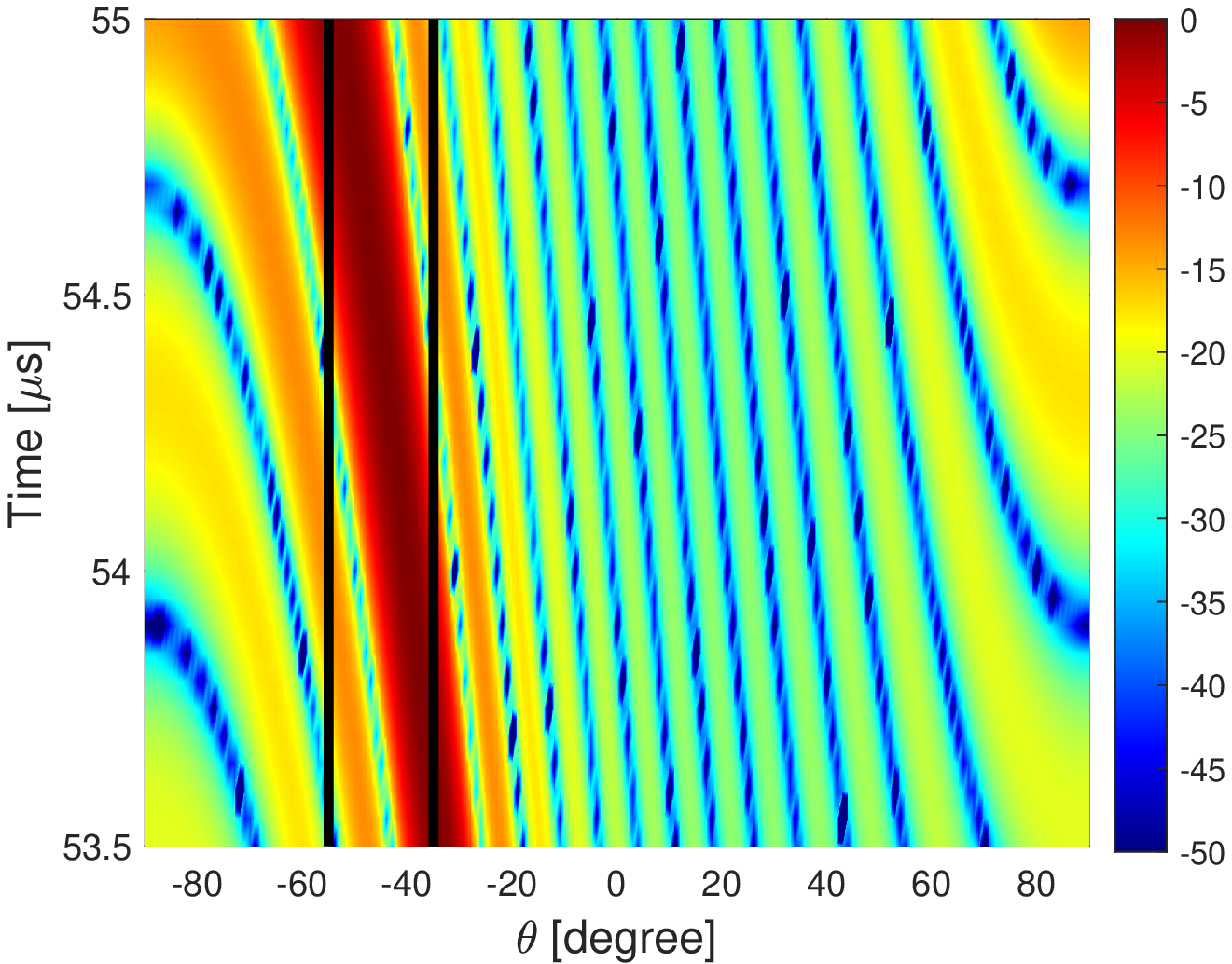}}
	\caption{Energy distribution along the azimuth dimension at a far-field distance of $r=15 \kern2pt km$.}
	\label{fig4}
\end{figure}

Evidenced by the derivation in \eqref{eq15}, the beam scanning is nonlinear with time.
Fig. \ref{fig4} shows the energy distribution of the FDA along the azimuth dimension for the same duration ($1.5 \kern 2pt \mu s$) at the far-field distance $r=15 \kern2pt km$, where the resulting scan time $t\in \left[ \frac{r}{c},\frac{r}{c}+{{T}_{p}} \right]$ in Fig. \ref{fig4a} and Fig. \ref{fig4b} are $t\in \left( 50,51.5 \right)\mu s$ and $t\in \left( 53.5,55 \right)\mu s$, respectively.
Here, a uniform weight vector has been used, with the FO being set to $\Delta f=80 \kern 2pt kHz$.
In the figure, the black straight lines mark the mainbeam extension. 
One may note from Fig. \ref{fig4} that, despite having the same duration, Fig. \ref{fig4b} exhibits a larger beam coverage than Fig. \ref{fig4a}. 
To be more precise, the beam coverage is about $14^o$ in Fig. \ref{fig4a}, and  $20^o$ in Fig. \ref{fig4b}.
The reason for this is that wider beam coverage results in faster scans over time, a result that is consistent with the derivation in \eqref{eq16}.

\subsubsection{Initial mainlobe direction}
\begin{figure}[ht]
	\centering  
	\subfigure[$\mathbf{w}={{\mathbf{1}}_{M}}$.]{
		\label{fig5a}
		\includegraphics[width=0.45\textwidth]{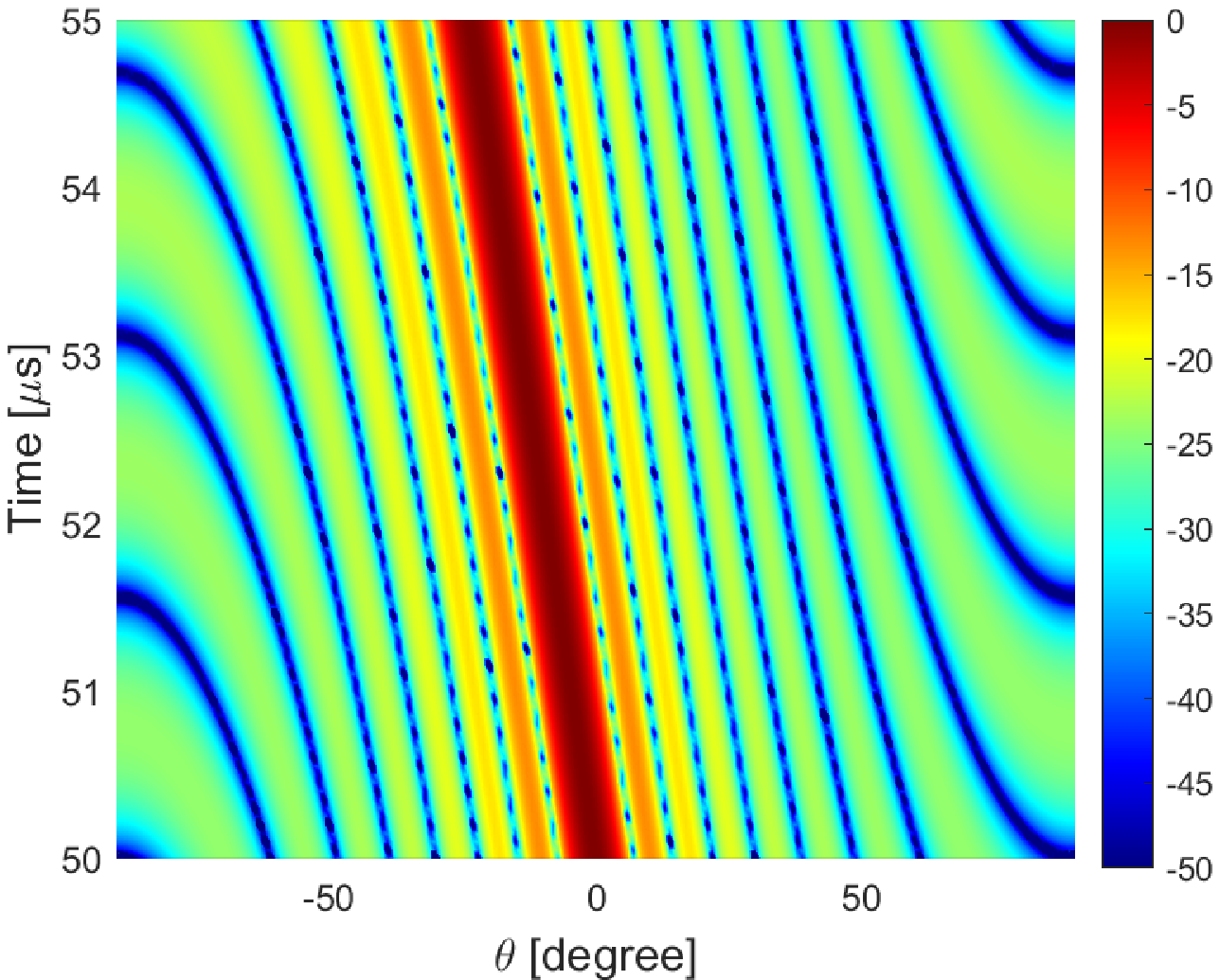}}
	\subfigure[$\mathbf{w}={{\left[ {{\mathbf{a}}_{T}}\left( {{60}^{o}} \right) \odot {{\mathbf{a}}_{T}}\left( \Delta f,{{60}^{o}} \right) \right]}^{c}}$.]{
		\label{fig5b}
		\includegraphics[width=0.45\textwidth]{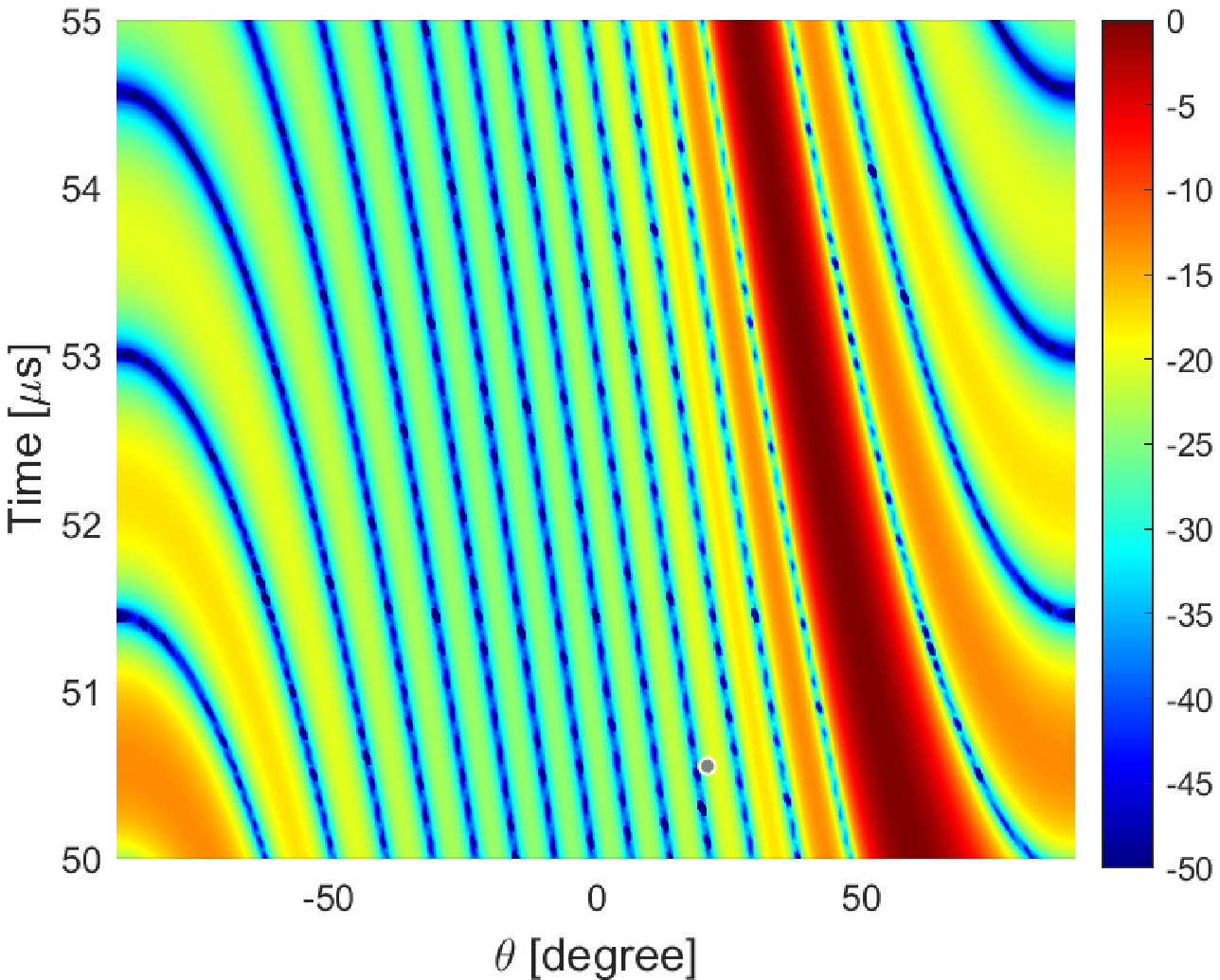}}
	\caption{The beampatterns for different transmit weight vectors.}
	\label{fig5}
\end{figure}

As shown, a uniform transmit weight vector causes the beam to scan from an azimuth of $0^o$.
Fig. \ref{fig5} illustrates the resulting beampatterns for different weight vectors, where the FO is $\Delta f=40 \kern 2pt kHz$.
Here, the transmit weight vector is set to $\mathbf{w}={{\left[ {{\mathbf{a}}_{T}}\left( {{0}^{o}} \right)*{{\mathbf{a}}_{T}}\left( \Delta f,{{0}^{o}} \right) \right]}^{c}}={{\mathbf{1}}_{M}}$ in Fig. \ref{fig5a}, and to $\mathbf{w}={{\left[ {{\mathbf{a}}_{T}}\left( {{60}^{o}} \right)*{{\mathbf{a}}_{T}}\left( \Delta f,{{60}^{o}} \right) \right]}^{c}}$ in Fig. \ref{fig5b}. 
As expected, the mainbeam starts scanning from the $ - {0^o }$ and ${60^o }$, respectively, indicating that the simulation results are consistent with results discussed in Section \ref{Sec3c}.

\subsubsection{Beampatterns at different distances}
\begin{figure}[ht]
	\centering  
	\subfigure[FITB at $r=27 \kern 2pt km$.]{
		\label{fig6a}
		\includegraphics[width=0.45\textwidth]{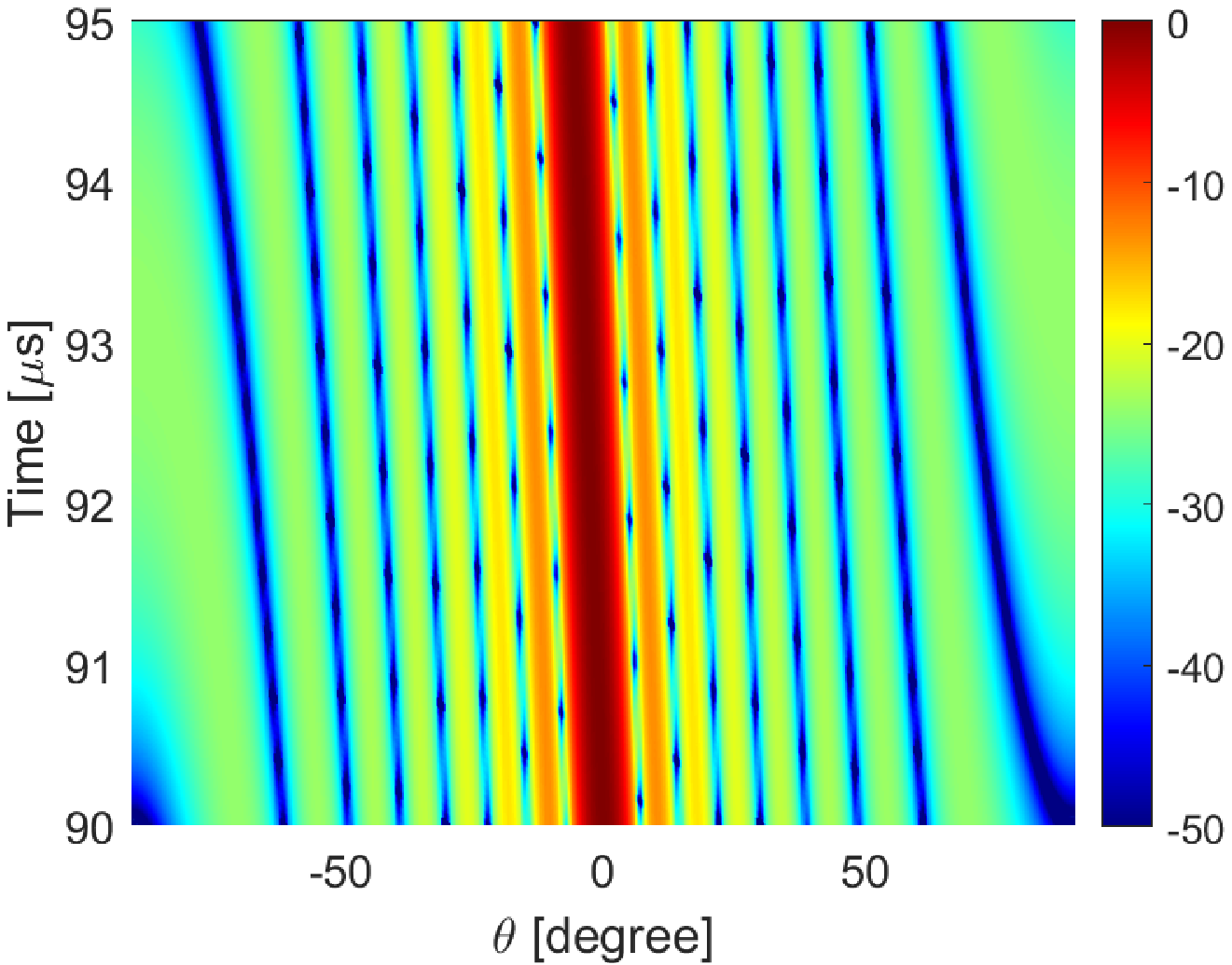}}
	\subfigure[Array factor \eqref{eq7} at $r=27 \kern 2pt km$.]{
		\label{fig6b}
		\includegraphics[width=0.45\textwidth]{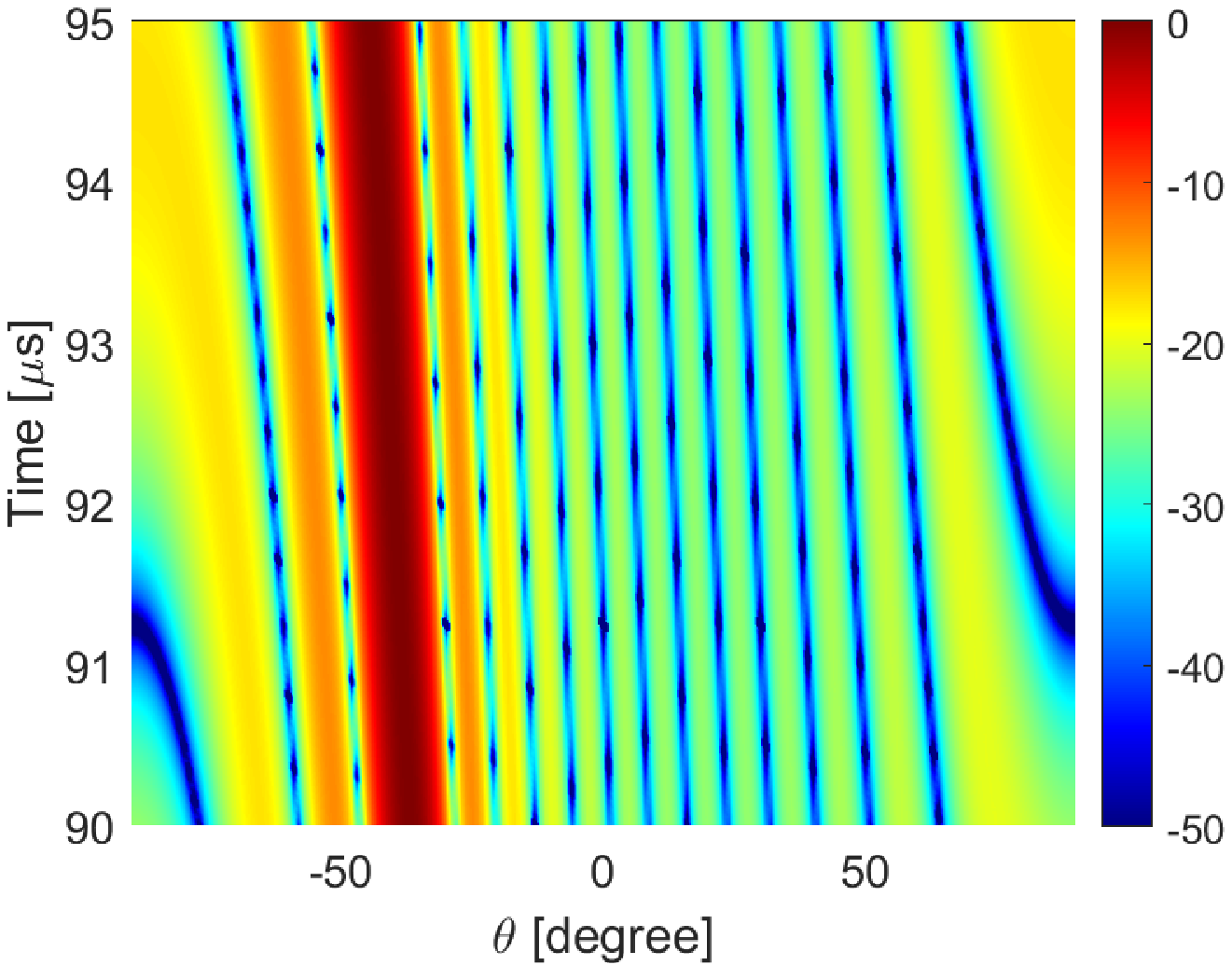}}
	\subfigure[FITB at $r=18 \kern 2pt km$.]{
		\label{fig6c}
		\includegraphics[width=0.45\textwidth]{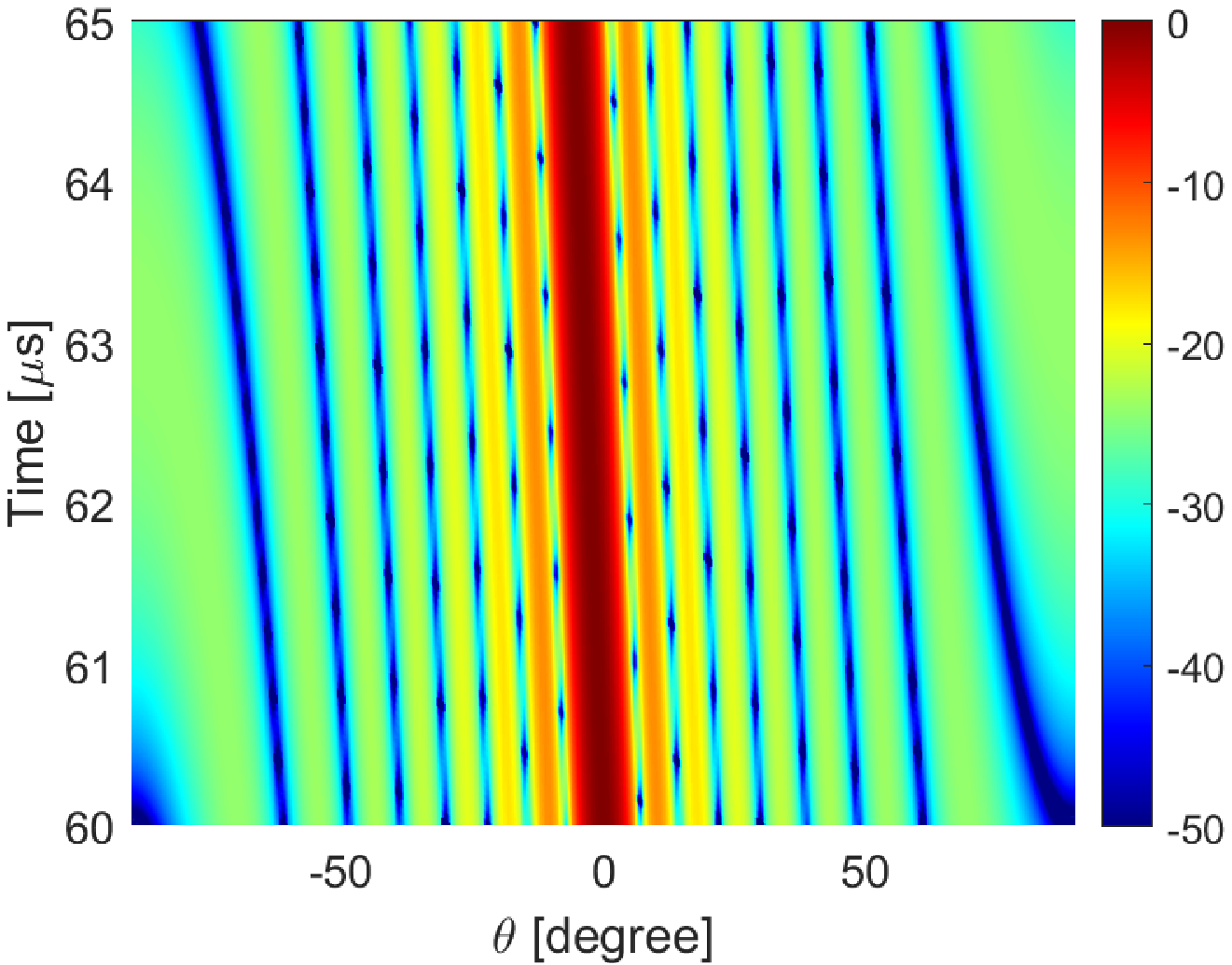}}
	\subfigure[Array factor \eqref{eq7} at $r=18 \kern 2pt km$.]{
		\label{fig6d}
		\includegraphics[width=0.45\textwidth]{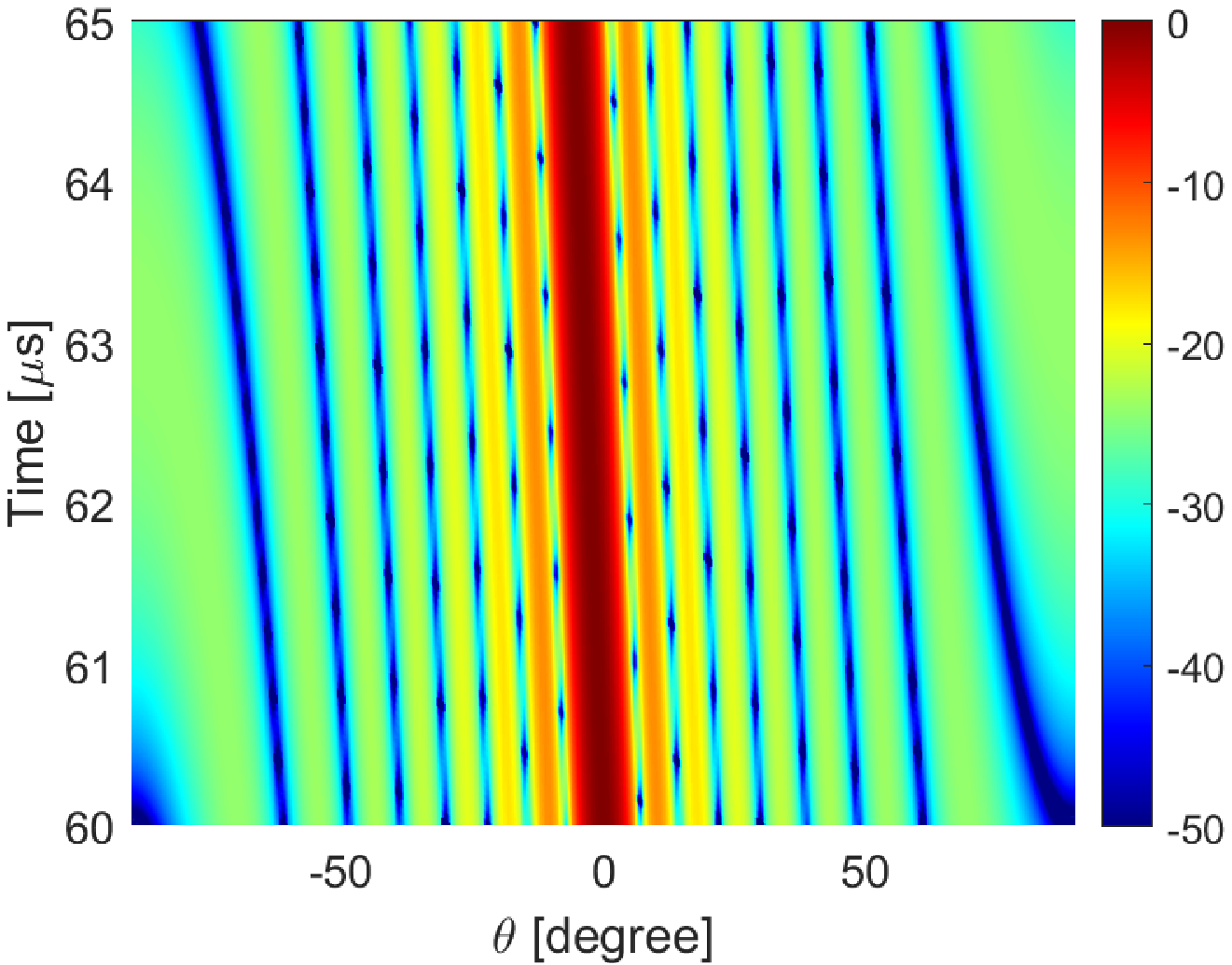}}
	\caption{The beampattern formed by the FDA array factors in \eqref{eq7} and the FITB in \eqref{eq9}, at distance $r=18 \kern 2pt km$ and $r=27 \kern 2pt km$.}
	\label{fig6}
\end{figure}

Existing studies using the array factor represented by \eqref{eq7} blur the relationship between time and range, holding that the FDA transmit beampattern is not only a function of time, but also a function of the target distance.
However, as our proposed FITB in \eqref{eq9} shows, it is angle-time-dependent but not range-dependent.
Fig. \ref{fig6} compares the beampatterns formed by the FDA array factors in \eqref{eq7} and the proposed FITB given in \eqref{eq9}, for the case when the same transmitted pulse travels to a range of $18\kern 2pt km$ and $27\kern 2pt km$, respectively, with the FO set to $\Delta f=10 \kern 2pt kHz$.
Comparing Figs. \ref{fig6a} and \ref{fig6c}, the simulation results of Figs. \ref{fig6b} and \ref{fig6d} reveal that when the same pulse propagates to different distances in free space, the direction of the mainlobe of the beam formed by the FDA changes.
This phenomenon clearly violates the law of electromagnetic wave propagation.
Furthermore, as can be seen from Figs. \ref{fig6a} and \ref{fig6c}, for a given distance in free space, the FDA beam automatically scans along the azimuth dimension for the duration of the pulse. Then, the identical auto-scanning process propagates forward at the speed of light.
In conclusion, our proposed FITB model is deemed to be more accurately representing the actual FDA beam transmission process than the one presented in \eqref{eq7}. 

\section{FDA and phased-array}\label{Sec4}
Considering the range-time-angle-dependent array factor in \eqref{eq7}, most conventional studies typically concentrate on designing FOs striving to decouple the time-variant transmit beampattern \cite{2014Subarray,2016Frequency}. 
A significant transmission gain can then be obtained by focusing the energy 
on the desired target in the range-angle space.
However, by analyzing the relationship between distance and time, it can be seen that the FDA transmit beam has auto-scanning properties. In this sector, the influence of FO on the proposed FITB and the relationship between the FDA and the phased-array will be discussed.

\subsection{Different FOs}
As the beampattern formed by the linear FO can be expressed in closed-form, its characteristics is possible to analyze in detail, with even nonlinear FOs, such as logarithmic \cite{6951408} FO and random \cite{2017The} FO, having been investigated in the FDA literature.
Consider the nonlinear FOs $\Delta {f_m},m = 0,1,...,M - 1$, for which the electric field can be expressed as
\begin{equation}
	\begin{aligned}
		& {{E}_{T}}\left( \theta ,t' \right)=\frac{1}{r}e\left( \theta \left| {{f}_{c}} \right. \right)\phi \left( {{t}'} \right){{e}^{j2\pi {{f}_{c}}{t}'}} \\ 
		& \times \underbrace{\sum\limits_{m=0}^{M-1}{{{e}^{j2\pi \Delta {{f}_{m}}{t}'}}{{e}^{j2\pi \left( {{f}_{c}}+\Delta {{f}_{m}} \right)\frac{md\sin \theta }{c}}}}}_{Array \kern 2pt Factor} \\ 
	\end{aligned}
\end{equation}
for ${t}'\in \left[ 0,{{T}_{p}} \right]$.
Then, the FITB can be expressed as
\begin{equation}
	{{A}_{nonlinear}}\left( t',\theta  \right)=\mathbf{a}_{T,nonlinear}^{T}\left( \theta  \right){{a}_{T,nonlinear}}\left( {{t}'} \right)
\end{equation}
where 
\begin{subequations}
	\begin{equation}
		\kern -33pt {{\mathbf{a}}_{T,nonlinear}}\left( {{t}'} \right)=\left[ \begin{matrix}
			{{e}^{j2\pi \Delta {{f}_{0}}{t}'}}  \\
			{{e}^{j2\pi \Delta {{f}_{1}}{t}'}}  \\
			...  \\
			{{e}^{j2\pi \Delta {{f}_{M-1}}{t}'}}  \\
		\end{matrix} \right]
	\end{equation}
	\begin{equation}
		\kern 24pt \mathbf{a}_{T,nonlinear}^{T}\left( \theta  \right)=\left[ \begin{matrix}
			{{e}^{j2\pi \left( {{f}_{c}}+\Delta {{f}_{0}} \right)\frac{d\sin \theta }{c}}}  \\
			{{e}^{j2\pi \left( {{f}_{c}}+\Delta {{f}_{1}} \right)\frac{d\sin \theta }{c}}}  \\
			...  \\
			{{e}^{j2\pi \left( {{f}_{c}}+\Delta {{f}_{M-1}} \right)\frac{d\sin \theta }{c}}}  \\
		\end{matrix} \right].
	\end{equation}
\end{subequations}
Similarly, for time-dependent or time-modulated FOs \cite{2014Frequency,shao2016time,yao2016solutions,yao2016synthesis}.
the FITB has the form
\begin{equation}
	{{A}_{T,time}}\left( t,\theta  \right)=\mathbf{a}_{T}^{T}\left( \theta  \right){{\mathbf{a}}_{T,time}}\left( {{t}'} \right)
\end{equation}
for $t' \in \left[ {0,{T_p}} \right]$, where
\begin{equation}
	{{\mathbf{a}}_{T,time}}\left( {{t}'} \right)=\left[ \begin{matrix}
		{{e}^{j2\pi {{\chi }_{0}}\left( {{t}'} \right){t}'}}  \\
		{{e}^{j2\pi {{\chi }_{1}}\left( {t}'+\frac{d\sin \theta }{c} \right)\left( {t}'+\frac{d\sin \theta }{c} \right)}}  \\
		...  \\
		{{e}^{j2\pi {{\chi }_{M-1}}\left( {t}'+\frac{\left( M-1 \right)d\sin \theta }{c} \right)\left( {t}'+\frac{\left( M-1 \right)d\sin \theta }{c} \right)}}  \\
	\end{matrix} \right],
\end{equation}
where ${{\chi }_{m}}\left( t' \right),m=0,1,...,M-1$, is a function of $t'$, such as $\sqrt[2]{\left( \centerdot  \right)}$,  $\sqrt[3]{\left( \centerdot  \right)}$, ${{\tan }^{-1}}\left( \centerdot  \right)$, or $\sinh \left( \centerdot  \right)$ \cite{yao2016solutions,yao2016synthesis}.

\subsection{FDA and phased-array}
In general, FDA can be regarded as a phased-array with time-variant weight vector.
If the weight vector 
\begin{equation}
	\mathbf{w}={{\left[ 1,{{e}^{j2\pi \Delta ft}},...,{{e}^{j2\pi \left( M-1 \right)\Delta ft}} \right]}^{T}},
\end{equation}
an auto-scanning transmit beam with zero initial direction will be generated.
Applying a time-variant weight vector in FDA, the FITB can be approximated as
\begin{equation}\label{eq31}
	{{A}_{\varphi }}\left( t',\theta  \right)\approx \left| \frac{\sin \left[ M\pi \left( \Delta f{t}'+\varphi \left( {{t}'} \right)+{{f}_{c}}\frac{d\sin \theta }{c} \right) \right]}{\sin \left[ \pi \left( \Delta f{t}'+\varphi \left( {{t}'} \right)+{{f}_{c}}\frac{d\sin \theta }{c} \right) \right]} \right|
\end{equation}
for $t' \in \left[ {0,{T_p}} \right]$,
where $\varphi \left( t' \right)$ is a slowly varying phase function.
The function in \eqref{eq31} has peaks at locations satisfying
\begin{equation}
	\Delta ft' + \varphi \left( t' \right) + \frac{{{f_c}d\sin \theta }}{c} = k,
\end{equation}
for $k \in \mathbb{Z}$.
Note that any desired beam scanning scheme can be realized by appropriately designing $\varphi \left( t' \right)$.
For example, if one expect the beam to be steered to $\theta  \in \left[ {{\theta _1},{\theta _2}} \right]$ at $t' \in \left[ {0,{T_1}} \right]$ and to $\theta  \in \left[ {{\theta _3},{\theta _4}} \right]$ at $t' \in \left[ {{T_2},{T_p}} \right]$, one can for ${t}'\in \left[ 0,{{T}_{1}} \right]$ and $\theta \in \left[ {{\theta }_{1}},{{\theta }_{2}} \right]$, $\varphi \left( t' \right)$ set
\begin{equation}
	\varphi \left( {{t}'} \right)=-\Delta f{t}'-\frac{{{f}_{c}}}{c}d\sin \left( \frac{{{\theta }_{2}}-{{\theta }_{1}}}{{{T}_{1}}}{t}'+{{\theta }_{1}} \right).
\end{equation}
whereas for ${t}'\in \left[ {{T}_{2}},{{T}_{p}} \right]$ and $\theta \in \left[ {{\theta }_{3}},{{\theta }_{4}} \right]$, 
\begin{equation}
	\varphi \left( {{t}'} \right)=-\Delta f{t}'-\frac{{{f}_{c}}}{c}d\sin \left[ \frac{{{\theta }_{4}}-{{\theta }_{3}}}{{{T}_{p}}-{{T}_{2}}}{t}'+{{\theta }_{3}}-\frac{{{T}_{2}}\left( {{\theta }_{4}}-{{\theta }_{3}} \right)}{{{T}_{p}}-{{T}_{2}}} \right].
\end{equation}
At other times and angles, one sets
\begin{equation}
	\varphi \left( {{t}'} \right)=-\Delta f{t}'-\frac{{{f}_{c}}}{c}d\sin \left( {{\theta }_{2}} \right).
\end{equation}
Interestingly, the array system designed in this scheme is very similar to the traditional phased-array, while it has only a weak connection to FDA.
Generally, phased-arrays use fixed transmit weights to steer the beam in the desired direction, whereas the designed array uses time-variant weights within the desired steering interval.
This also illustrates how the FDA solution to achieve the desired beam steering makes the designed array system converge to the time-variant phased-array, suggesting that using a uniform FO is the optimal choice for the FDA transmit beampattern.
We also note that a uniform FO can be realized using direct digital synthesis (DDS) technology \cite{cordesses2004directtt}.

\subsection{Numerical Simulation}
\begin{figure}[t]
	\centering  
	\subfigure[]{
		\label{fig7a}
		\includegraphics[width=0.45\textwidth]{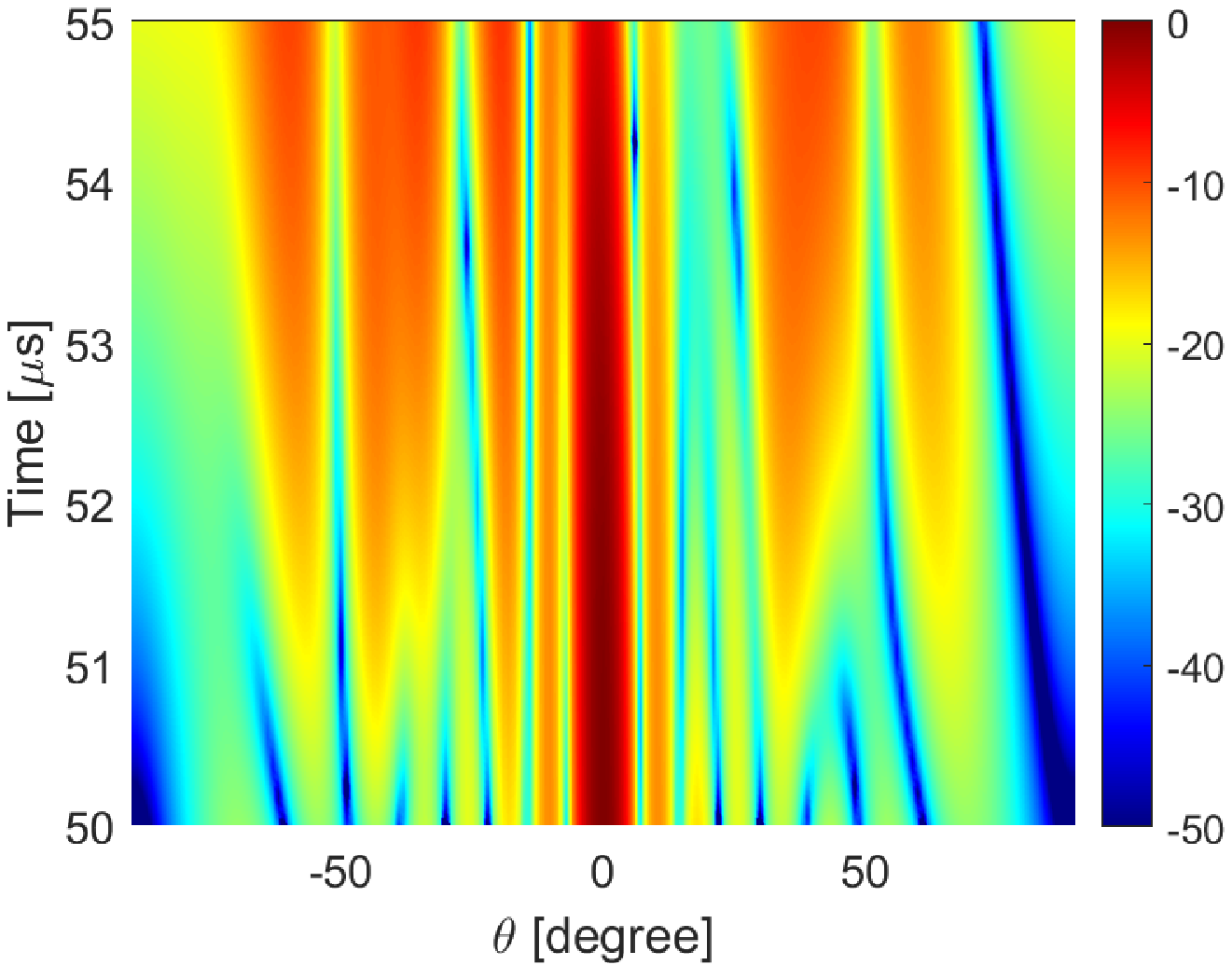}}
	\subfigure[]{
		\label{fig7b}
		\includegraphics[width=0.45\textwidth]{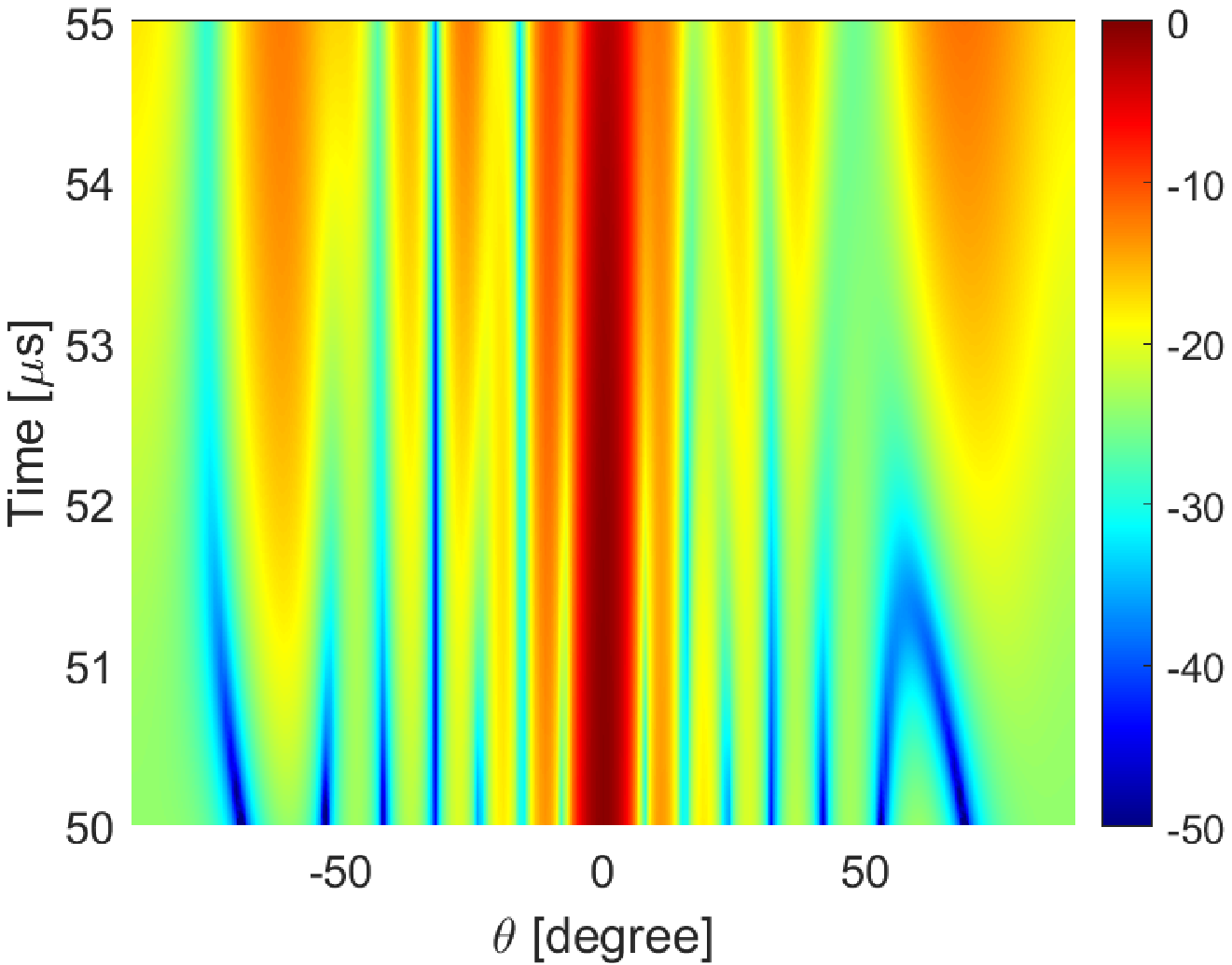}}
	\subfigure[]{
		\label{fig7c}
		\includegraphics[width=0.45\textwidth]{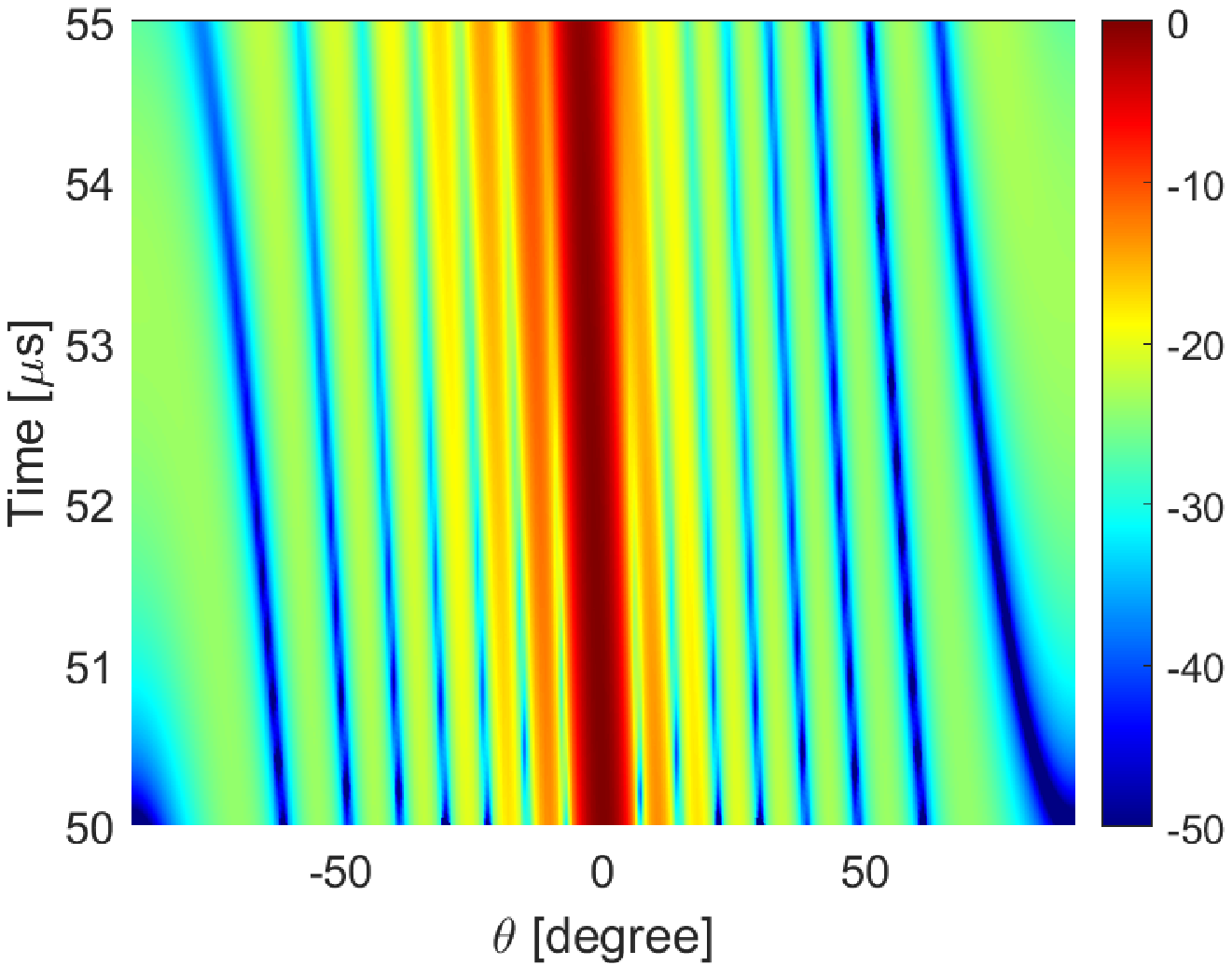}}
	\subfigure[]{
		\label{fig7d}
		\includegraphics[width=0.45\textwidth]{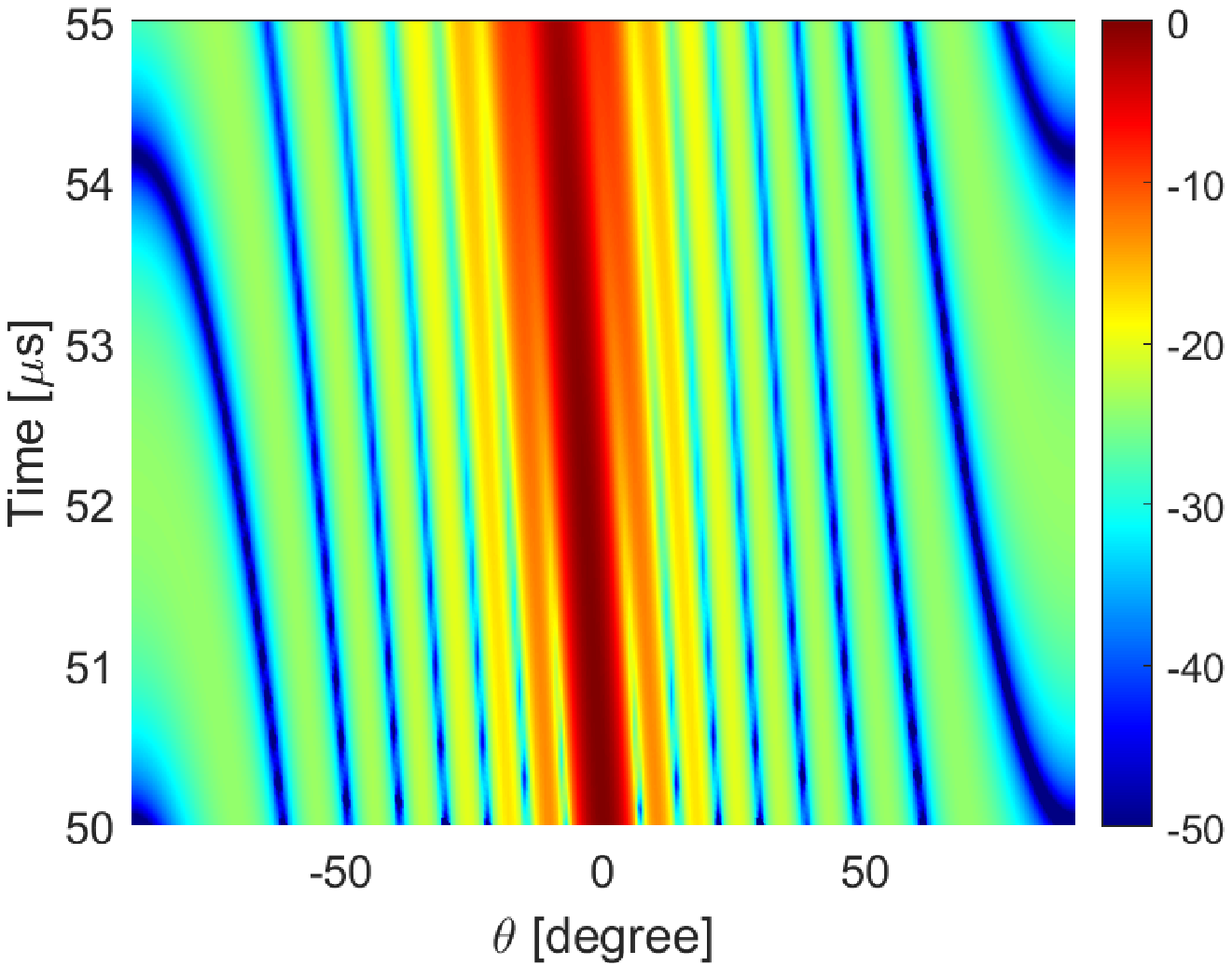}}
	\caption{The FITB at a given distance $r = 15 \kern 2pt km$ with four different types of FO coding methods: (a) random FOs $\Delta {{f}_{m}}={{\varepsilon }_{m}}\Delta f$, where $\Delta f=100 \kern 2pt kHz$ and ${{\varepsilon }_{m}}$ is a uniformly distributed random number in the interval $\left( 0,1 \right)$; (b) Costas-coded FOs $\Delta {{f}_{m}}={{\operatorname{c}}_{m}} \Delta f$, where $\Delta f=5 \kern 2pt kHz$ and ${{c}_{m}}$ denotes the Costas code; (c) logarithmic FOs $\Delta {f_m} = \ln \left( {m + 1} \right)\Delta f$, where $\Delta f =  50\kern 2ptkHz$; (d) square FOs $\Delta {f_m} = {m^2}\Delta f$, where $\Delta f =  1\kern 2ptkHz$, for $m=0,1,...,M-1$.}
	\label{fig7}
\end{figure}

We proceed to consider the FDA simulation parameters detailed in Section \ref{Sec3d}.
Fig. \ref{fig7} depicts the FITB for different types of FOs, including the random \cite{2017The}, Costas-coded \cite{wang2016range}, logarithmic \cite{6951408}, and square FOs \cite{yao2016synthesis}.
The FOs used above are widely used to design the FDA transmit beampattern focused in the range-angele domain.
It may be seen that whether configured with nonlinear FOs or time-modulated FOs, the resulting FDA always produces a beampattern with its mainlobe focusing in the initial direction (in these simulation, the uniform weight vector results in the beam steering to azimuth $0^o$), similar to the static beampattern of a phased-array.
This is due to the fact that the FDA using these FOs will generate a focused beampattern without considering the range-time relationship. Therefore, the range-time relationship will appear to be static (approximatively).

\section{FDA integral transmit beampattern}\label{Sec5}
As the proposed FITB is time-dependent, it may be viewed as obtaining all the energy radiated to a certain azimuth during the pulse duration, i.e., as a FDA integral transmit beampattern (FGTB), defined as
\begin{equation}\label{eq36}
	\begin{aligned}
		& {{A}_{\text{FGTB}}}\left( \mathbf{s}\left( {{t}'} \right),\mathbf{w},\Delta f,\theta  \right)= \\ 
		& =\frac{1}{{{T}_{p}}}\int\limits_{0}^{{{T}_{p}}}{\left\{ \begin{aligned}
				& \mathbf{s}\left( {{t}'} \right){{\mathbf{s}}^{H}}\left( {{t}'} \right){{\mathbf{w}}^{H}} \\ 
				& {{\mathbf{a}}_{T}}\left( {t}',\Delta f,\theta  \right)\mathbf{a}_{T}^{H}\left( {t}',\Delta f,\theta  \right)\mathbf{w} \\ 
			\end{aligned} \right\}\operatorname{d}{t}'} \\ 
		& =\frac{1}{{{T}_{p}}}\operatorname{Tr}\left\{ {{\mathbf{R}}_{\text{FDA}}}\left[ \mathbf{w}\odot {{{\mathbf{\bar{a}}}}_{T}}\left( \Delta f,\theta  \right) \right]{{\left[ \mathbf{w}\odot {{{\mathbf{\bar{a}}}}_{T}}\left( \Delta f,\theta  \right) \right]}^{H}} \right\} \\ 
	\end{aligned}
\end{equation}
where 
\begin{subequations}
	\begin{equation}
		{{\mathbf{a}}_{T}}\left( {t}',\Delta f,\theta  \right)={{\mathbf{a}}_{T}}\left( {{t}'} \right)\odot {{\mathbf{a}}_{T}}\left( \theta  \right)\odot {{\mathbf{a}}_{T}}\left( \Delta f,\theta  \right)
	\end{equation}
	\begin{equation}
		\kern -24pt {{{\mathbf{\bar{a}}}}_{T}}\left( \Delta f,\theta  \right)={{\mathbf{a}}_{T}}\left( \theta  \right)\odot {{\mathbf{a}}_{T}}\left( \Delta f,\theta  \right),
	\end{equation}
\end{subequations}
and
\begin{equation}
	{{\mathbf{R}}_{\text{FDA}}}=\int\limits_{0}^{{{T}_{p}}}{\left\{ \mathbf{s}\left( {{t}'} \right){{\mathbf{a}}_{T}}\left( {{t}'} \right)\mathbf{a}_{T}^{H}\left( {{t}'} \right){{\mathbf{s}}^{H}}\left( t \right) \right\}\operatorname{d}{t}'},
\end{equation}
denotes the covariance matrix of the FDA transmitted signal, with $\operatorname{Tr}\left\{ \cdot  \right\}$ being the trace operator, and
\begin{equation}
	\mathbf{s}\left( t' \right)={{\left[ {{s}_{0}}\left( t \right),{{s}_{1}}\left( t \right),...,{{s}_{M-1}}\left( t \right) \right]}^{T}}
\end{equation}
is the baseband waveform vector, with
${{s}_{m}}\left( t \right),m=1,...,M-1$, denoting the baseband envelope of the $m$-th array element.
This implies that the FGTB can be considered as the result of integrating the proposed FITB over the entire pulse duration.
For a co-located MIMO array, the FGTB transmit beampattern \cite{li2007mimo,liii2008mimo} can be expressed as
\begin{equation}\label{eq40}
	{{A}_{\text{MIMO}}}\left( \mathbf{w},\theta  \right)=\operatorname{Tr}\left\{ {{\mathbf{R}}_{\text{MIMO}}}\left[ \mathbf{w}\odot {{\mathbf{a}}_{T}}\left( \theta  \right) \right]{{\left[ \mathbf{w}\odot {{\mathbf{a}}_{T}}\left( \theta  \right) \right]}^{H}} \right\}
\end{equation}
where ${{\mathbf{R}}_{\text{MIMO}}}=\int\limits_{0}^{{{T}_{p}}}{\mathbf{s}\left( {{t}'} \right){{\mathbf{s}}^{H}}\left( t' \right)\operatorname{d}{t}'}\in {{\mathbb{C}}^{M\times M}}$ denotes the covariance matrix of the MIMO waveform vector $\mathbf{s}\left( {{t}'} \right)$.
It is worth noting that the derived FGTB has a form consistent with the conventional MIMO radar beampattern.
The only difference is in the steering vectors ${{\mathbf{\bar{a}}}_{T}}\left( \Delta f,\theta  \right)={{\mathbf{a}}_{T}}\left( \theta  \right)\odot {{\mathbf{a}}_{T}}\left( \Delta f,\theta  \right)$ and ${{\mathbf{a}}_{T}}\left( \theta  \right)$.
Consider that the MIMO and FDA transmit signals can be expressed as
\begin{equation}
	{{u}_{MIMO}}\left( t \right)=\sum\limits_{m=1}^{{{N}_{T}}}{{{s}_{m}}\left( t \right){{e}^{j2\pi {{f}_{c}}t}}}
\end{equation}
and
\begin{equation}
	{{u}_{FDA}}\left( t \right)=\sum\limits_{m=1}^{{{N}_{T}}}{\left[ {{s}_{m}}\left( t \right){{e}^{j2\pi \left( m-1 \right)\Delta ft}} \right]{{e}^{j2\pi {{f}_{c}}t}}},
\end{equation}
respectively. Therefore, if the contribution of the FO-dependent steering vector ${{\mathbf{a}}_{T}}\left( \Delta f,\theta  \right)$ can be neglected in FGTB, the problem of FGTB synthesis based on waveform optimization is equivalent to the beampattern design of co-located MIMO. 
Then, by replacing the designed waveform equivalently with the FDA transmit waveform, massive MIMO transmit beampattern design techniques (see, e.g., \cite{stoica2007probing,ahmed2014mimo,zhang2015mimo,aubry2016mimo}) can be applied to the FGTB synthesis in parallel.
Notably, if
\begin{equation}
	2\pi \Delta f\frac{{{m^2}d\sin \theta }}{c} \leqslant \frac{\pi }{4},
\end{equation}
for $ m = 0,1,...,M - 1$, the contribution of the vector ${{\mathbf{a}}_{T}}\left( \Delta f,\theta  \right)$ to the FGTB can be neglected.
Combined with the narrowband assumption in \eqref{eq3}, the FO is required to satisfy the condition 
\begin{equation}\label{eq44}
	\frac{M\cdot {{B}_{{{s}_{m}}\left( {{t}'} \right)}}-{{f}_{c}}}{2M}\le \Delta f\le \frac{{{f}_{c}}}{4{{M}^{2}}-M}
\end{equation}
for $ m = 0,1,...,M - 1$.
Therefore, the FO-dependent steering vector ${{\mathbf{a}}_{T}}\left( \Delta f,\theta  \right)$ should be taken into account for scenarios with large array apertures and high range resolution requirements.
In particular, this holds for FDA-MIMO systems \cite{sammartino2013frequency,7181636},
as their configuration requires that each element transmits the baseband waveforms with non-overlapping spectra, i.e., $\Delta f\ge {{B}_{{{s}_{m}}\left( t' \right)}}$.
In this case, the FOs will not meet the requirements in \eqref{eq44}, and FO effects must be considered.

It is interesting to examine two extreme cases, for which the properties of the FGTB can be explained analytically.
For $M\cdot \Delta f\ll {{B}_{{{s}_{m}}\left( t \right)}},m=0,1,...,M-1$, ${{\mathbf{R}}_\text{FDA}} \approx {\mathbf{1}_{M \times M}}$, the FGTB can be approximately expressed as
\begin{equation}
	{{A}_{\text{FGTB}}}\left( \mathbf{w},\Delta f,\theta  \right)\approx \frac{1}{{{T}_{p}}}{{\left| {{\mathbf{w}}^{H}}{{{\mathbf{\bar{a}}}}_{T}}\left( \Delta f,\theta  \right) \right|}^{2}},
\end{equation} 
which is thus similar to the beampattern of a coherent MIMO array.
For $\Delta f\geqslant {{B}_{{{s}_{m}}\left( t \right)}},m=0,1,...,M-1$, ${{\mathbf{R}}_\text{FDA}} \approx {\mathbf{I}_{M}}$ with ${\mathbf{I}}$ denoting the identity matrix,
the FGTB can be approximated as 
\begin{equation}
	{A_\text{FGTB}}\left( {{\mathbf{w}} } \right) \approx \frac{1}{{{T_p}}}{\left\| {\mathbf{w}} \right\|^2}
\end{equation} 
where $\left\|  \cdot  \right\|$ denotes the ${l_2}$ norm, indicating that the FGTB acts in a omnidirectional manner, corresponding to an orthogonal MIMO beampattern.

\subsection{Numerical Simulation}
\subsubsection{FGTB}
\begin{figure}
	\centering
	\includegraphics[width=0.6\textwidth]{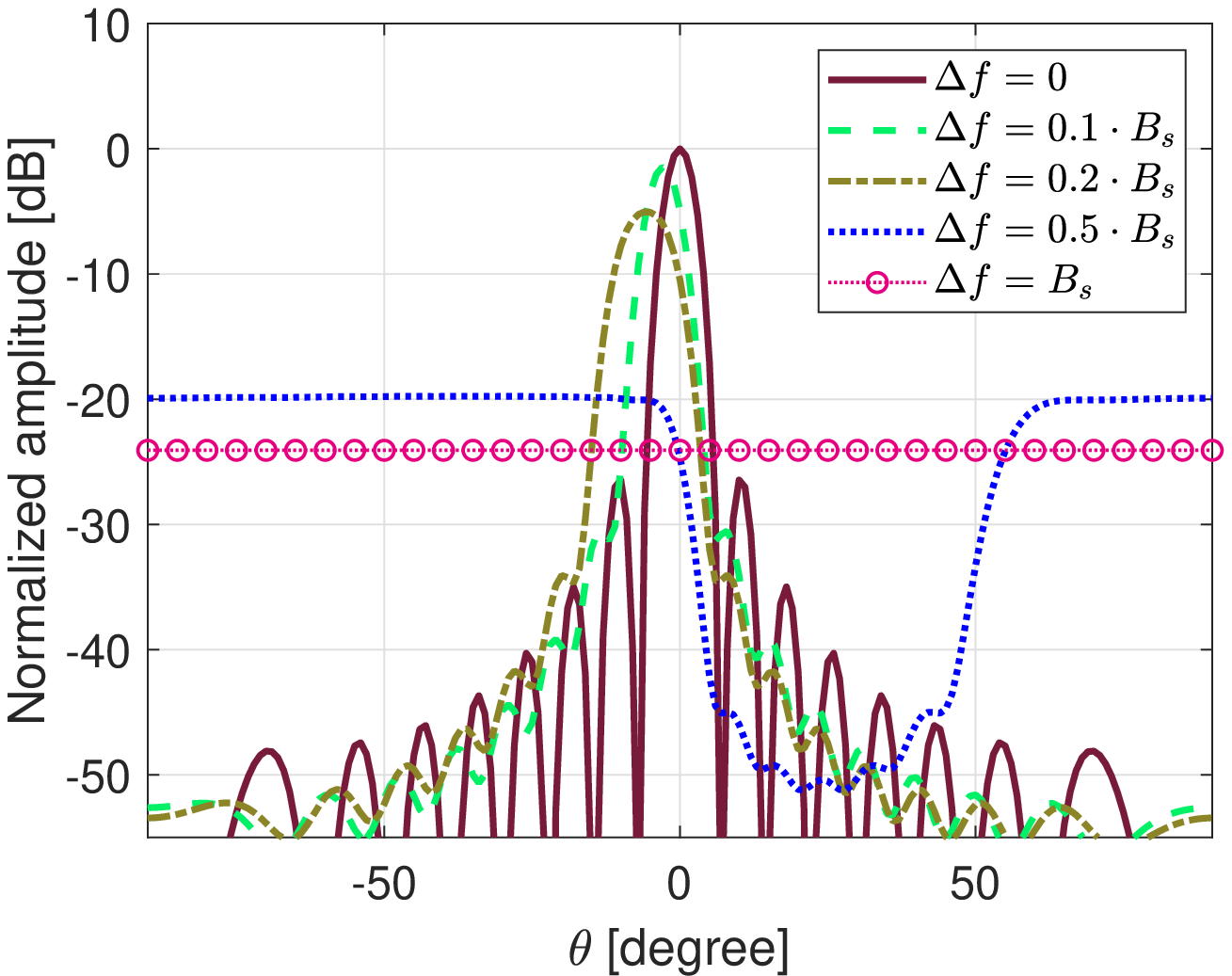}
	\caption{The illustration of FDA integral transmit beampattern.}\label{fig8}
\end{figure}

Fig. \ref{fig8} depicts the FGTB, where the bandwidth of the baseband waveform transmitted by each array element is ${{B}_{s}}=10 \kern 2pt MHz$, with the remaining simulation parameters being same as in Section \ref{Sec3d}.
It is noted that as the FO increases, the coherent gain decreases, whereas the mainlobe width of the beam grows.
For $\Delta f = {B_s}$, the transmitted waveforms are orthogonal, resulting in an omnidirectional energy radiation, such that the coherent gain does not exist.
These results can also be explained from the perspective of beam scanning as analyzed in Section \ref{Sec3d}.
For small FOs, the beam coverage is negligible during the pulse duration, so the FGTB is similar to the conventional coherent MIMO array.
That is, there will be a clear peak at azimuth $0^o$.
When $\Delta f\ge {{B}_{{{s}_{m}}\left( t' \right)}}$, the FDA scans though the whole visible azimuth sector within one pulse duration, which corresponds to the omnidirectional FGTB in \eqref{eq44}.

\subsubsection{The influence of the FO on the FGTB}
\begin{figure}[t]
	\centering  
	\subfigure[Uniform
	transmit weight vector ${\mathbf{w}} = {{\mathbf{1}}_M}$.]{
		\label{fig9a}
		\includegraphics[width=0.6\textwidth]{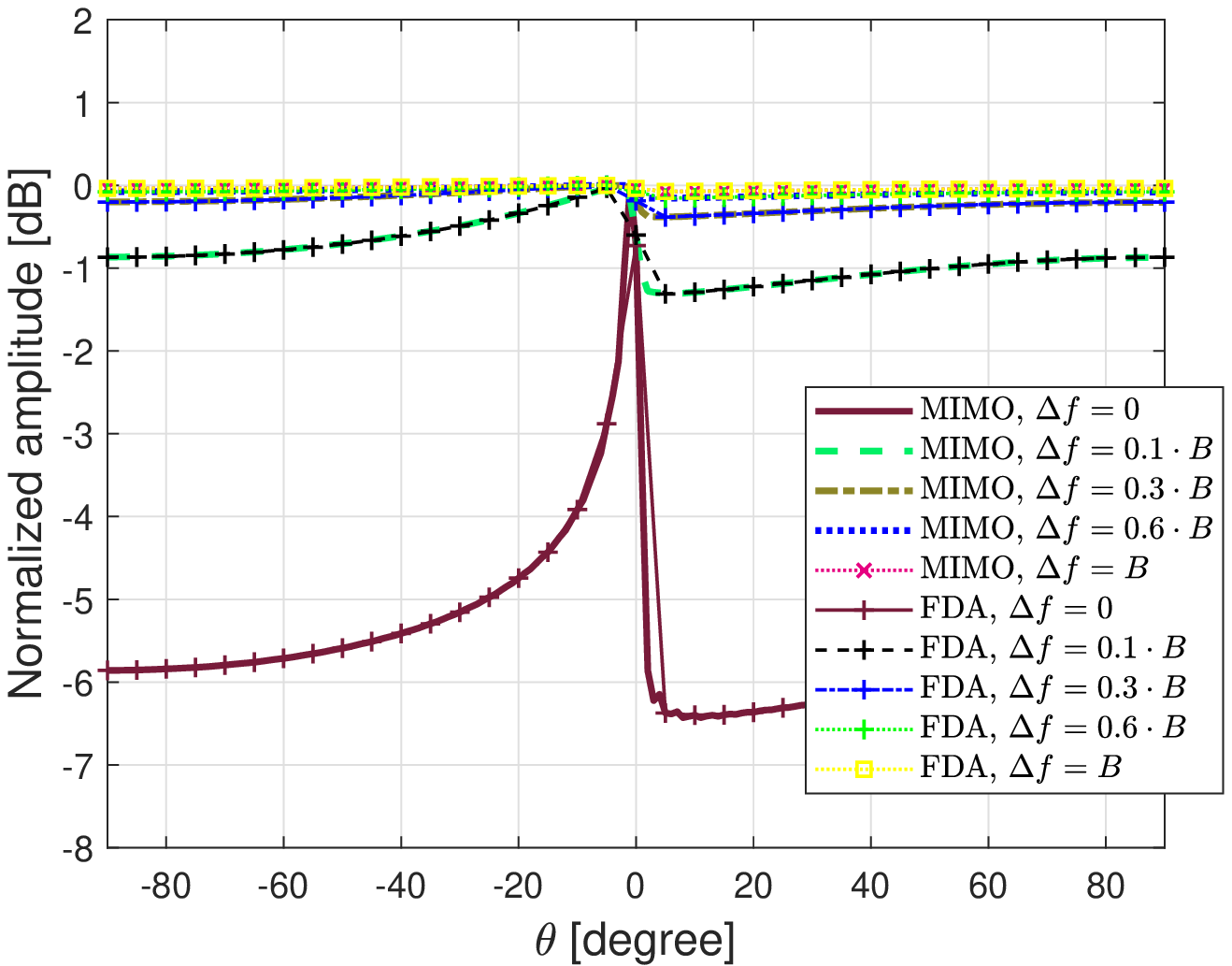}}
	\subfigure[Random transmit weight vector ${w_m} = {e^{j2\pi {c_m}}}$, for $m = 1,...,M - 1$, where ${c_m}$ is a uniformly distributed random number in the interval $\left( 0,1 \right)$.]{
		\label{fig9b}
		\includegraphics[width=0.6\textwidth]{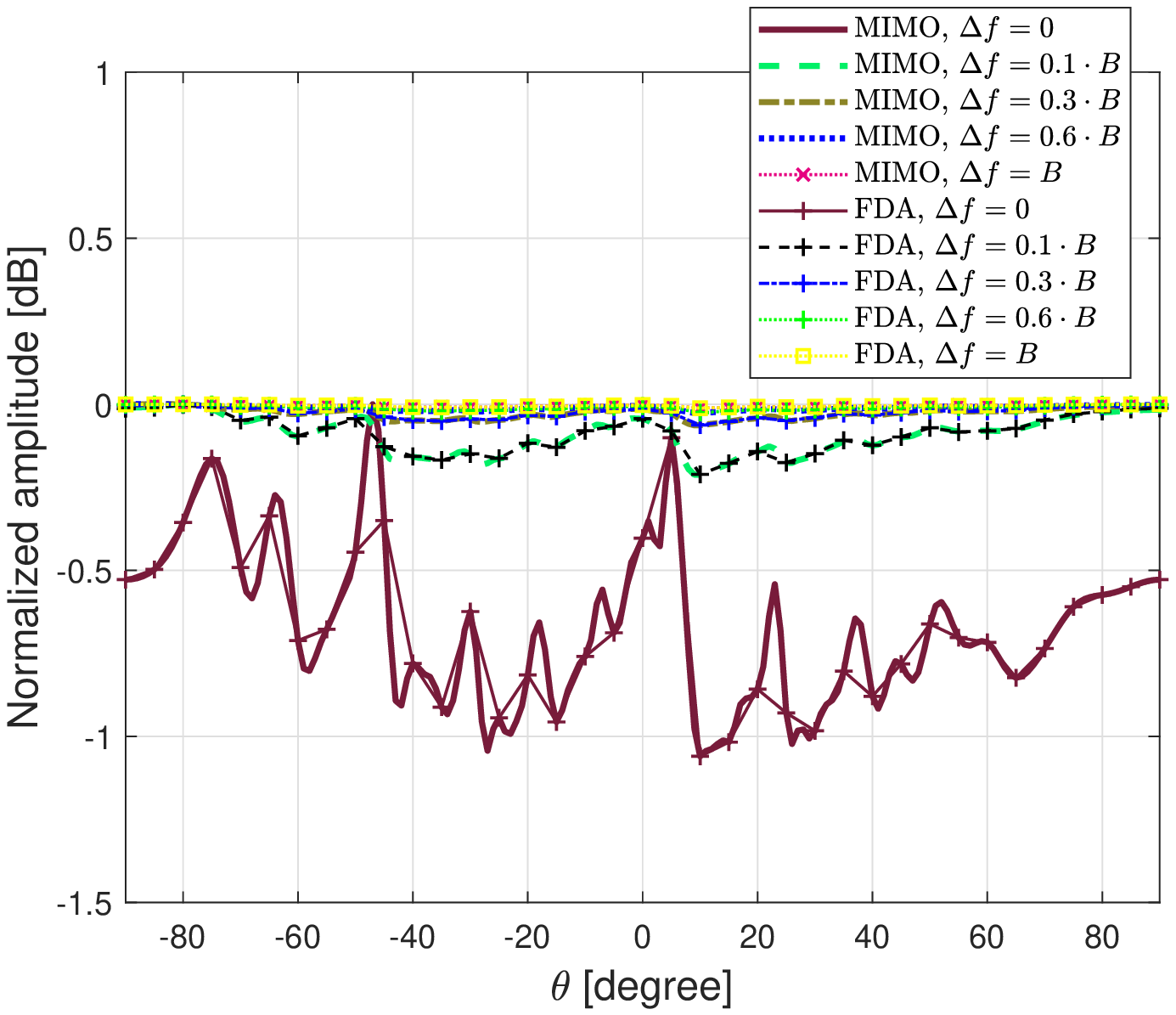}}
	\caption{Comparison of the FGTB and co-located MIMO transmit beampattern.}
	\label{fig9}
\end{figure}

To achieve the transmit waveform analogy for a co-located MIMO, a chirp signal with an additional uniform FOs is depolyed.
Then, the basedband waveform transmitted by the $m$-th element of the MIMO array can be written as
\begin{subequations}
	\begin{equation}
		{{s}_{m,MIMO}}\left( t \right)=\operatorname{rect}\left( \frac{t}{{{T}_{p}}} \right){{e}^{j\pi {{\gamma }_{m}}{{t}^{2}}}}{{e}^{j2\pi m\Delta ft}}
	\end{equation}
\end{subequations}
where 
\begin{equation}
	\operatorname{rect}\left( x \right)=\left\{ \begin{matrix}
		1 & \left| x \right|\le \frac{1}{2}  \\
		0 & \left| x \right|>\frac{1}{2}  \\
	\end{matrix} \right.
\end{equation}
and ${{\gamma }_{m}}=\frac{100+10\cdot m}{{{T}_{p}}}$, for $m=0,1,...M-1$, denotes the chirp rate of the $m$th waveform.
For FDA, the $m$-th baseband waveform is
\begin{equation}
	{{s}_{m,FDA}}\left( t \right)=\operatorname{rect}\left( \frac{t}{{{T}_{p}}} \right){{e}^{j\pi {{\gamma }_{m}}{{t}^{2}}}}.
\end{equation}
It should be emphasized that, for MIMO, the beampattern is calculated based on \eqref{eq40}, while for FDA, the calculation is based on \eqref{eq36}. 
As an example, consider a case with $M = 40$ transmit array elements, with all other simulation parameters are as given in \ref{Sec3d}.
Fig. \ref{fig9} shows the simulation of the FGTB and the co-located MIMO beampatterns, where the reference bandwidth is $B=10{\kern 2pt}MHz$.
As can be seen in Fig. \ref{fig9a}, in the case of uniform weights and a small FO, the beampattern still has a clear peak at azimuth $0^o$.
The increase in FO causes the coherence of the transmitted waveform to decrease, resulting in a gradually flat beampattern and the disappearance of peaks.
As shown in Fig. \ref{fig9b}, this condition exists regardless of the transmit weights.
Most importantly, it can still be clearly observed that there is no obvious difference between the FGTB and MIMO beampatterns regardless of FOs being small or large (the curves overlap).
This shows that theoretically, although the large FO does not satisfy the equivalent conditions in \eqref{eq44}, the two beampatterns are equivalent.
In fact, this reveals that the FDA can be regarded as a MIMO radar for which the phase change is caused by the FO.
However, the resulting phase has a negligible effect on the energy distribution at a given location in free space, as expected for the proposed FGTB. At the same time, if the energy distribution of the whole space is considered, the presence of FOs causes the beam to automatically scan, which corresponds to the proposed FITB.

\section{Conclusion}\label{Sec6}
In this paper, the time-range characteristics of the FDA transmit beampattern has been detailed.
Two FDA transmit beampattern models, namely, the instantaneous transmit beampattern and the integral transmit beampattern, have been formulated and analyzed. 
Numerical simulations show that the presence of a small FO gives the FDA the ability to transmit scanning beams automatically.
Furthermore, we show that the often made assumption of range-angle-time dependent FDA transmit beampattern is not accurate. 
In addition, it is shown that the FO-dependent steering vector constitutes the only difference between the proposed FGTB and the MIMO beampattern, creating a better understanding of the FDA range-time relations.
\newline

\bibliographystyle{unsrt}
\bibliography{Refs}

\end{document}